%% file: A1_near_field_fringes.tex
\documentclass[sn-mathphys,iicol]{sn-jnl} 

\usepackage{graphicx}
\usepackage{dcolumn}
\usepackage{bm}
\usepackage{hyperref}
\usepackage[separate-uncertainty = true]{siunitx}
\usepackage{placeins}
\usepackage{lineno}
\usepackage{ulem}
\usepackage[thinc]{esdiff}

\usepackage{xcolor}
\definecolor{black_red}{rgb}{0.00,0.00,0.00} 

\bibstyle{sn-mathphys}

\begin{document}

\title[ ]{Attosecond physics in optical near fields}

\author*{\fnm{Jonas} \sur{Heimerl$^1$}}\email{jonas.heimerl@fau.de}
\affil{$^1$\orgdiv{Department of Physics}, \orgname{Friedrich-Alexander-Universität Erlangen-Nürnberg (FAU)}, \orgaddress{\street{Staudtstraße 1}, \city{Erlangen}, \postcode{91058}, \country{Germany}}}
\equalcont{These authors contributed equally to this work.}

\author{\fnm{Stefan} \sur{Meier$^1$}}
\equalcont{These authors contributed equally to this work.}

\author{\fnm{Anne} \sur{Herzig$^2$}}%
\affil{$^2$\orgdiv{Institute of Physics and Department of Life, Light and Matter}, \orgname{University of Rostock}, \orgaddress{ \city{Rostock}, \postcode{18051}, \country{Germany}}}
\author{\fnm{Felix} \sur{López Hoffmann$^1$}}%
\author{\fnm{Lennart} \sur{Seiffert$^2$}}%
\author{\fnm{Daniel} \sur{Lesko$^1$}}%
\author{\fnm{Simon} \sur{Hillmann$^1$}}%
\author{\fnm{Simon} \sur{Wittigschlager$^1$}}%
\author{\fnm{Tobias} \sur{Weitz$^1$}}
\author{\fnm{Thomas} \sur{Fennel$^2$}}%
\author{\fnm{Peter} \sur{Hommelhoff$^{1,3}$}}
\affil{$^3$\orgdiv{Faculty of Physics}, \orgname{Ludwig-Maximilians-Universität München}, \orgaddress{\city{München}, \postcode{80799}, \country{Germany}}}

\date{\today}

\abstract{ \bf{
Attosecond science, the electron control by the field of ultrashort laser pulses, is maturing into lightfield-driven electronics, called petahertz electronics~\cite{Krausz2014, Heide2024}. Based on optical field-driven nanostructures, elements for petahertz electronics have been demonstrated~\cite{Schiffrin2012,Rybka2016,Karnetzky2018,Ludwig2019,Bionta2021,Pettine2024}. These hinge on the understanding of the electron dynamics in the optical near field of the nanostructure~\cite{Krger2011,Herink2012, Park2012,Dombi2013,Thomas2013,Hobbs2014,Schtz2018,Krger2018,Blochl2022,Kim2023,Dienstbier2023}. 
Here we show near field-induced low energy stripes (NILES) in carrier-envelope phase-dependent electron spectra, a new spectral feature appearing in the direct electrons emitted from a strongly driven nanostructure, i.e., in the easily accessible energy region between 0 and a few electron volts. 
NILES emerge due to the sub-cycle sensitivity of ponderomotive acceleration of electrons injected into a strong near field gradient by a few-cycle optical waveform. 
NILES enables us to track the emission of direct and re-scattered electrons down to sub-cycle time-scales and to infer the electron momentum width at emission. Because NILES shows up in the direct part of the electrons, a large fraction of the emitted electrons can now be steered in new ways, facilitating the isolation of individual electron bursts with high charge density of 430 attosecond duration. These results not only substantially advance the understanding of attosecond physics in optical near fields, but also provide new ways of electron control for the nascent field of petahertz electronics.
}
}

\maketitle
Measuring the energy of electrons or photons emitted from a sample under laser irradiation leads to electron or photon spectra. They contain tell-tale features such as multi-photon orders~\cite{Agostini1979,Paulus1994}, the plateau of high-harmonic generation and rescattering physics~\cite{Ferray1988,Corkum1993, Schafer1993,Paulus1994, PaulusBeckerNicklichEtAl1994}, the featureless cut-off indicating the generation of individual attosecond pulses \cite{Hentschel2001}, or the low-energy structure~\cite{Blaga2009, Quan2009}, evidencing soft electron recollision mechanisms \cite{Liu2010, Kstner2012,Lemell2012}. Discovering and understanding these qualitatively new spectral features facilitated new dimensions of insights, giving birth to strong field and attosecond science~\cite{Ferray1988} and often providing new control knobs to the motion of electron wavepackets on sub-optical cycle timescales~\cite{Krger2011, Kim2023, Dienstbier2023}.

Most of these spectral features have been measured with atoms and molecules in the gas phase, but also at nanoparticles~\cite{Zherebtsov2011, Seiffert2022} or nanometer sharp metal structures~\cite{Schenk2010,Krger2011,Herink2012,Piglosiewicz2013,Dombi2013,Bionta2013}. In contrast to atoms or molecules, nanometric objects like spheres or needle tips generate an optical near field when illuminated with light. This near field typically leads to a substantial field enhancement at the surface of the structure and, from there, decays away into the vacuum. Hence, a dramatic optical field inhomogeneity arises, making the optical near field a temporally {\it and} spatially sharply varying field. But it is this near field in which electrons emitted from the nanostructure move. It thus has a pivotal influence on the electron motion, which leads to, for example, the quenching of the electron's quiver motion~\cite{Herink2012,Echternkamp2016,Schtz2018}. Resting on these localized and enhanced optical near fields, metal nanostructures are today the basis for initial attosecond-fast on-chip devices. They enable sub-optical cycle field sampling and light-driven current generation
~\cite{Schiffrin2012,Rybka2016,Karnetzky2018,Keathley2019,Ludwig2019,Bionta2021,Pettine2024} as well as ultrafast tunneling microscopes~\cite{Yoshioka2018,Garg2019}.

The intense optical near field leads to a ponderomotive potential, and the fast spatial decay to a steep ponderomotive potential gradient, exerting a force on a photo-emitted electron. We will show that a photo-emitted electron experiences an additional drift velocity from the quickly (spatially) varying ponderomotive potential that continuously decreases the later the electron is born into the laser pulse. Vice versa, especially electrons born early into the optical field {\it always} gain energy, even if they are born into the field at the crest of the optical field cycle. This is in contrast to most cases known from atomic physics where these electrons would not show any drift momentum - here they do, resulting in a new {\it minimum energy curve}. Together with the temporal localization of the electron time of birth in the tunneling regime and the few cycle nature of the driving field, near field-induced low energy stripes result in the direct part of the electron spectrum, as we will show below in detail.  Interestingly, this goes far beyond the quenching of the quiver motion~\cite{Herink2012}, which also shows up for long driving pulses, for example. NILES do not.

\section*{Near field-induced low energy stripes, NILES}

\begin{figure*}
    \centering
    \includegraphics[width=0.7\linewidth]{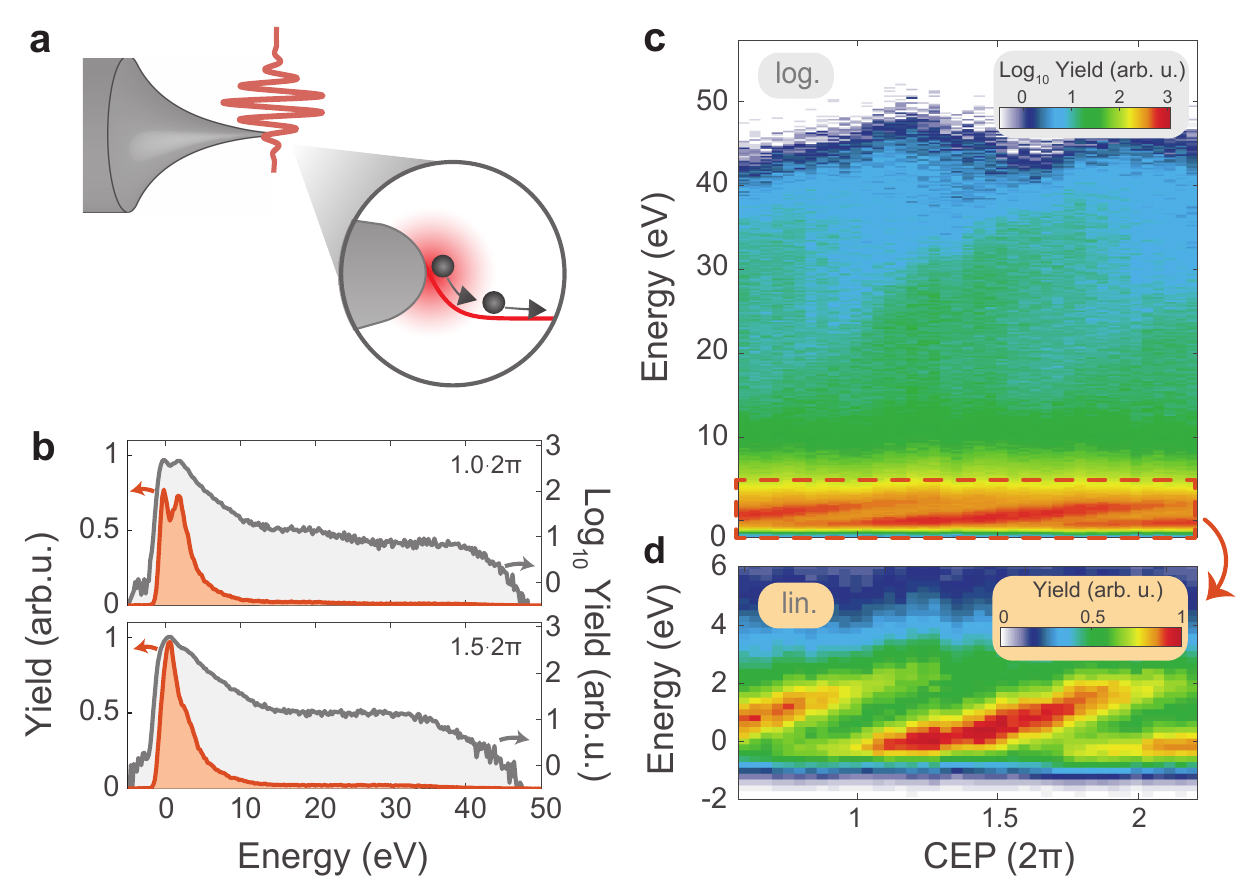}
    \caption{\textbf{Near field-induced low energy stripes, NILES}:
    \textbf{a}~Electrons are photo-emitted by a two-cycle laser pulse and propagate in the steeply decaying ponderomotive potential resulting from the optical near field (red in zoom-in) at the apex of a sharp tungsten tip. With just $\sim$2 emission events per laser pulse, this leads to NILES (see text). \textbf{b}~Measured electron energy spectra for two selected carrier-envelope phases (as indicated) shown on linear scale (orange, left axis) and  logarithmic scale (gray, right axis). \textbf{c}~CEP-resolved spectra. The high energy electrons show the well-known CEP-dependent energy modulation in the plateau ($\sim$10 to 40\,eV) and the cut-off ($\sim$40 to 50 eV). In the low-energy region between $-2$\,eV and 5\,eV new spectral stripes appear prominently: NILES. \textbf{d}~Zoom-in on the low-energy region of direct electrons showing NILES, on a linear yield scale. The large modulation depth of the integrated yield with CEP, reaching up to $33\,\%$ is noteworthy (see Methods). Lineouts of the low-energy region are shown in Extended Data Fig.~\ref{fig:lineouts_experiment}. The energy of 30\,eV resulting from the static voltage of $-30$\,V applied to the tip is subtracted from the energy axis in c and d.
    }
    \label{fig:experiment_NILES}
\end{figure*}

We trigger electrons from a sharp tungsten needle tip with \SI{10}{\nano\meter} radius of curvature illuminated with $11.5$\,fs long (2.2 cycles) laser pulses at 1570\,nm (Fig.~\ref{fig:experiment_NILES}a; see Methods). The near field's peak intensity of \SI{2.0e13}{\watt\per\square\centi\meter} is chosen such that we are well in the strong field regime (Keldysh $\gamma \sim 0.7$)~\cite{Schenk2010}. This is substantiated by the clearly visible hallmarks of recollision physics (cut-offs of direct and recollision electrons and a recollision plateau) in the measured electron energy spectra, shown for two selected carrier-envelope phases (CEP) in Fig.~\ref{fig:experiment_NILES}b. The energy of \SI{30}{\eV} resulting from the static bias voltage applied to the tip is subtracted.

Intriguingly, CEP-resolved spectra in Fig.~\ref{fig:experiment_NILES}c reveal a pronounced CEP dependence over the entire spectrum, not only in the plateau and cut-off region but also for the direct electrons, i.e., below 5\,eV (shown on linear scale in Fig.~\ref{fig:experiment_NILES}d). We observe stripes that extend over a CEP of $\sim 1.4\times (2\pi)$, i.e., over more than one period. This low energy region showing NILES contains  51\% of all emitted electrons (range from -1\,eV to 4\,eV). When only one single peak is visible (CEP = $1.65 \times(2\pi)$), the energy width is $\sim 2.2$\,eV (full-width at half maximum, see also Extended Data Fig.~\ref{fig:lineouts_experiment}). The integrated yield in an energy window of $\pm 0.4$\,eV around the maximum shows a modulation depth of 33\% (see Methods), i.e., the total current modulation is about an order of magnitude higher than what is known from cut-off modulation-dominated works (see, e.g., \cite{Krger2011}).

\begin{figure*}
    \centering
    \includegraphics[width=0.99\linewidth]{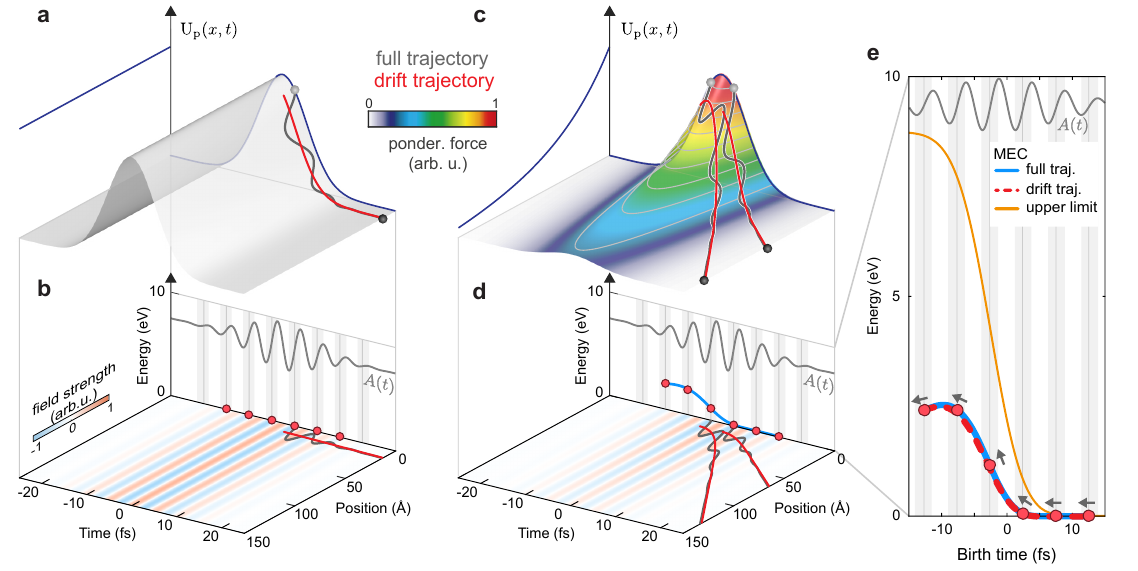}
    \caption{\textbf{Origin of NILES}: \textbf{a,c}~Position- and time-dependent ponderomotive potential $U_\mathrm{P}(x,t)$ of (a) a homogeneous optical field profile and (c) an optical near field with decay length $\zeta=\SI{10}{nm}$ as in the experiment. The projections (blue curves) indicate the respective spatial enhancement profiles and temporal intensity envelopes. The spatial gradient of this potential $\mathrm{d}U_\mathrm{P}(x,t)/\mathrm{d}x$ leads to a force acting on emitted electrons (surface color). Red curves show  trajectories of electrons liberated at instants of vanishing vector potential in the dominant field half cycles and resulting from the ponderomotive (cycle-averaged) force acting on the electrons. These trajectories are called drift trajectories, resulting in the {\it drift momentum}. The grey curves, by contrast, show the corresponding trajectories obtained when the full field is taken into account. Importantly, they result in the same final momentum, the drift momentum. \textbf{b,d}~Projections of the trajectories in position and time. The lowest possible drift momentum and respective {\it minimum final energy} is realized for electrons liberated at vector potentials $A(t)=0$ and becomes zero for a vanishing ponderomotive force, while additional ponderomotive acceleration results in {\it non-vanishing minimum energies} (compare red dots in b,d). The blue curve connecting the red spheres represents the minimum energy determined from full trajectories when varying the CEP. \textbf{e}~Zoom-in to panel d: For increasing CEP the minimum energies shift continuously as indicated by the arrows. This results in two almost identical minimum energy curves for the full trajectories and the drift trajectories (compare blue and dashed red curves). Clearly, the drift trajectories contain already the full physics. Importantly, the separated stripes of NILES result from the fact that the red dots are {\it sparse} on this curve and hence display a large energy difference exceeding $\sim\SI{1.2}{eV}$. The orange curve shows the analytical upper estimate for the minimum energy curve (MEC) as detailed in the main text. }
    \label{fig:explanation_NILES_Rostock}
\end{figure*}

The root cause for NILES is most straightforwardly understood when focusing on a comparison between electrons moving in a {\it quickly decaying} ponderomotive potential, like in front of a sharp nanostructure, and in a {\it spatially homogeneous} ponderomotive potential, such as when the electrons are emitted from individual atoms or molecules, or from a metallic structure with a large radius of curvature. In Fig.~\ref{fig:explanation_NILES_Rostock}a, we show the ponderomotive potential $U_\mathrm{P}(x,t)$ of an intense, short and spatially {\it homogeneous} optical field as a function of position and time in one spatial dimension, i.e., constant in space and with a Gaussian envelope in time (see Methods). Because we are in the optical tunneling emission regime, electrons are preferentially emitted at the crests of the negative half-cycles of the laser field. Hence, we first  concentrate on trajectories of these electrons.

For these electrons we can describe the action of the laser field twofold: (1)~If we consider the (cycle-resolved) optical field, the quiver motion of the electron eventually ceases and no net momentum is acquired because the vector potential $A(t) = - \int_{- \infty}^{t} E(t') dt'$ vanishes at the field crests. (2)~In the cycle-averaged intensity picture, the electron is born into a ponderomotive potential $U_\mathrm{P}(x,t)$ that is space-{\it independent} in the case of a homogeneous field. As its spatial gradient is $\nabla  U_\mathrm{P}(x,t) =0$, no force can act on the electron. The trajectories of both pictures are shown in Fig.~\ref{fig:explanation_NILES_Rostock}a-b, one including the quiver motion (full trajectories, case (1), grey curves), the other only including the drift motion in the ponderomotive potential picture (drift trajectories, case (2), red curves). Importantly, both pictures yield the same result: an electron emitted at the field crest exhibits no final kinetic energy. This repeats for every laser cycle (red dots in Fig.~\ref{fig:explanation_NILES_Rostock}b).

The situation at the sharp metal tip (Fig.~\ref{fig:explanation_NILES_Rostock}c-d) is entirely different because of the strong spatial decay of the optical near field. The field in front of the tip can be described by the position-dependent field enhancement factor $\xi(x)=1+(\xi_0-1)\cdot\exp(-x/\zeta)$, with the field enhancement factor $\xi_0$ at the surface of the tip, the distance from the tip apex \(x\), and the decay length $\zeta$, equaling the tip radius~\cite{Thomas2013,Seiffert2018}. The optical near field acting on the emitted electron is thus $E_\mathrm{NF}(x,t) = \xi(x) \cdot E_\mathrm{in}(t)$, with the incident laser field $E_\mathrm{in}(t)$ (see Methods for details). Consequently, the ponderomotive potential ($\propto \langle E^2_\mathrm{NF}(x,t)\rangle_t$) exhibits a large spatial gradient that photo-emitted electrons can roll down, cf. Fig.~\ref{fig:explanation_NILES_Rostock}c. This leads to non-zero drift momenta despite the trajectories being launched at vanishing vector potentials \cite{Kibble1966} and causes birth time-dependent final energies, see red dots in Fig.~\ref{fig:explanation_NILES_Rostock}d. Increasing the CEP and thus decreasing the birth times leads to a continuous shift of these dots (cf. arrows in Fig.~\ref{fig:explanation_NILES_Rostock}e). This forms what we call the {\it minimum energy curve} (MEC, blue curves in Figs.~\ref{fig:explanation_NILES_Rostock}d,e) because electrons emitted before or after the field crests gain more energy due to the non-vanishing vector potential. Most importantly, the MEC matches the behavior extracted from drift trajectories almost exactly (compare red dashed to blue curve in Fig.~\ref{fig:explanation_NILES_Rostock}e), substantiating the dominance of the ponderomotive acceleration over higher order effects of the quiver motion for NILES to appear.

In the following we provide an intuitive picture behind the MEC, which reflects a temporal integration of the spatio-temporal ponderomotive force along the numerically propagated trajectories. An analytical upper limit can be obtained when considering only the temporal evolution of the ponderomotive force sampled at the tip surface. For electrons starting at rest at the crests of the optical field cycles, this upper limit for the MEC reads (see Methods)
\begin{equation}
    \mathcal{E}_\mathrm{min.\,final}(t_b)  = \frac{2}{m}\left(U_\mathrm{P}^\mathrm{inc}\xi_0\xi_0'\right)^2 \mathcal{F}^2(t_b).
    \label{eq:MEC_analytical}
\end{equation}
Here, the electron energies are pivotally determined by the laser parameters (ponderomotive potential $U_\mathrm{P}^\mathrm{inc}$ of the incident field) as well as the magnitude $\xi_0$ and gradient $\xi'_0 \hspace{2mm} (= \mathrm{d}\xi(x)/\mathrm{d}x \vert_{x=0})$ of the near field enhancement at the tip surface. The birth time-dependent shape of the MEC is governed by the remaining normalized pulse fluence $\mathcal{F}(t_b) = \frac{1}{I_0}\int_{t_b}^\infty I(t) \mathrm{d}t$ associated with the intensity envelope $I(t)$ and experienced by the electron after its birth. Importantly, for a Gaussian $I(t)$, $\mathcal{F}(t_b)$ decreases continuously with $t_b$ in the shape of a complementary error function. The resulting analytical MEC (orange curve in Fig.~\ref{fig:explanation_NILES_Rostock}e) provides an upper limit as the ponderomotive force is maximal at the surface. MECs extracted from the trajectories show the same shape but with a smaller magnitude, which is because electrons leave the vicinity of the tip quickly for the experimental near field parameters. Importantly, the analytical upper bound represents a decent estimate as it predicts the order of magnitude of the minimal energies, the curve's shape and, most importantly, provides a direct link between the continuous increase of the minimal electron energy for decreasing (earlier) birth times and the pulse intensity envelope.

We emphasize that despite the conceptual equivalence of the considered ponderomotive acceleration to that reported for atoms~\cite{Bucksbaum1987, Agostini1987}, NILES is a specific feature of combined sub-wavelength field localization and few-cycle driving: For few-cycle pulses, only around two tunneling emission events arise during the interaction of the laser pulse with the tip, spaced in time by the period of the optical field. Because the duration of the few cycle laser pulses is so extremely short, the kinetic energy gained by the electrons (eq.~\ref{eq:MEC_analytical}) varies substantially between the emission events. The appearance of NILES thus results from the sparsity of the emission events within a few-cycle laser pulse together with the ponderomotive force. Intriguingly, when varying the CEP, the emission events continuously map out the MEC, resulting in the spectrally isolated stripe-like features, which enable sub-cycle resolution - the key novelty of NILES. We add in passing that while this effect is fully general, it is easier to observe with mid-IR driving wavelengths compared to visible or near-IR driving because of the $U_p^2$  scaling of the MEC (eq.~\ref{eq:MEC_analytical}, see Methods).

\begin{figure*}[ht]
\includegraphics[width=0.95\linewidth]{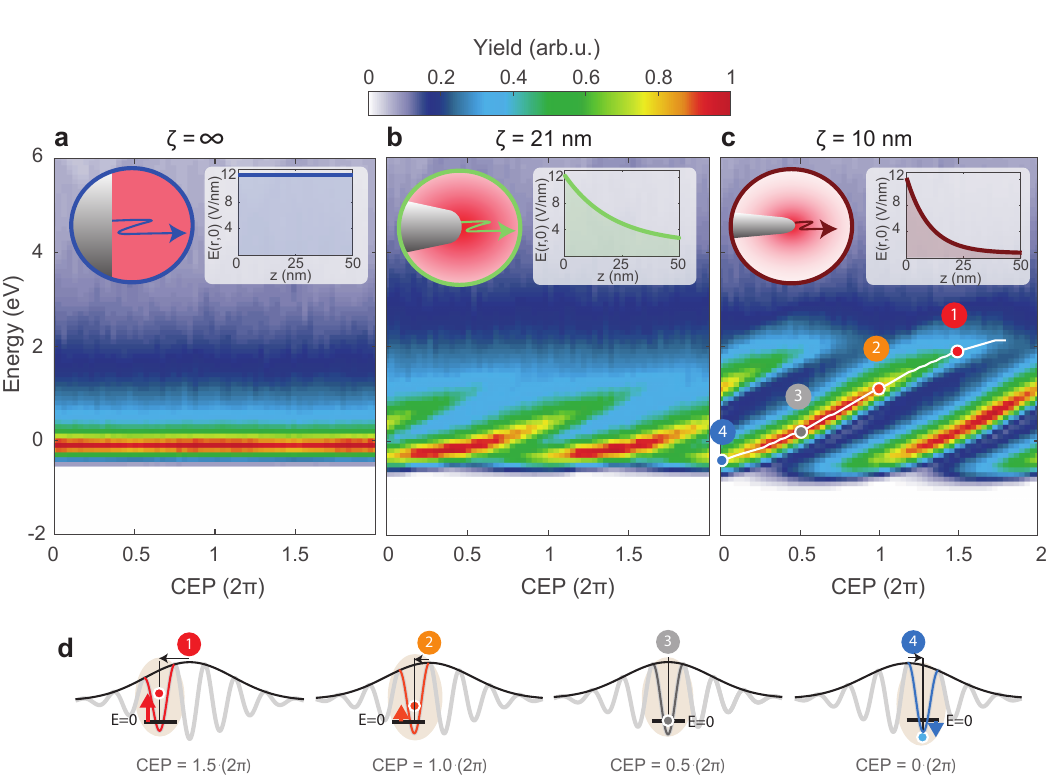}
\caption{\textbf{Decay length dependence of NILES}: \textbf{a}-\textbf{c}~Semi-classical simulations of electron spectra for three near field decay lengths as indicated above each panel and corresponding to (a) a planar surface, (b) a medium-sharp tip and (c) a sharp tip as used in the experiment. The near field peak intensity was fixed at $\SI{2e13}{W/cm^2}$. Clearly, NILES appear only for inhomogeneous near field profiles (b,c) and extend up to 2\,eV for the experimental parameters (c). The slope of NILES sampled in the central region increases from $0.76$\,eV/$(2\pi)$ in (b) to $1.5$\,eV/$(2\pi)$ in (c). We note that the slope scales indirectly proportional to the pulse duration, resulting in vanishing NILES features for long pulses (see Methods). The circles labeled 1 to 4 indicate different emission times within a selected half cycle. \textbf{d}~Tracking of the emission time by modulating the CEP. With respect to the homogeneous field case (energy \(E=0\), black line), emission before the center of the pulse (1, 2) leads to an energy up-shift,  while a later emission (4) results in an energy down-shift (indicated by the colored arrows). This behavior is given mainly by the MEC (cf.  Fig.~\ref{fig:explanation_NILES_Rostock}e); the shift to energies below $E=0$ is due to the fact that we now included the DC bias voltage applied to the tip (see Methods). The labels 1-4 relate directly to the corresponding one in panel c. We stress the excellent overall agreement of the numerical results in panel c with the experimental NILES data of Fig.~\ref{fig:experiment_NILES}. We have applied a Gaussian blurring filter of 0.3 eV (FWHM) to the semi-classical simulation results to account for the finite energy resolution of the detector in the experiment. As in Fig.~\ref{fig:experiment_NILES}, the energy resulting from the tip bias voltage is subtracted from the energy axis (see Methods).
\label{fig:niles-scheme}
}
\end{figure*}

To inspect the parameter dependencies of NILES we simulated CEP-resolved spectra for three different near field profiles as shown Fig.~\ref{fig:niles-scheme}: one spatially homogeneous ($\zeta \rightarrow \infty$, Fig.~\ref{fig:niles-scheme}a), one slowly decaying ($\zeta =  \SI{21}{\nano\meter}$, Fig.~\ref{fig:niles-scheme}b) and one fast decaying near field ($\zeta =  \SI{10}{\nano\meter}$, Fig.~\ref{fig:niles-scheme}c). Clearly, NILES show up for the two inhomogeneous field profiles (b,c), while NILES are absent if there is no spatial gradient of the near field-generated ponderomotive potential (a). As indicated by the four highlighted CEPs (circles 1-4), the continuous shifting of one relevant cycle can be observed over $\sim$$1.5$ CEP periods, i.e., even beyond one full period, similar to the experiment (Fig.\ref{fig:experiment_NILES}d).

Because of the strongly non-linear nature of the electron emission as function of field strength, only the central cycles of a laser pulse contribute substantially to the emission (see also Extended Data Fig.~\ref{fig:abstract_cycles}). For this reason, for any given CEP no more than two NILES are visible in the experimental spectra and no more than three in the simulation (for an optical pulse duration of $\tau = \SI{11.5}{\femto\second}$ at 1570\,nm), corresponding to the 2-3 mainly emitting optical cycles for the respective CEP.

The structure of NILES depends strongly on the pulse duration $\tau$, which determines the number of visible stripes. The slope of these stripes in the central region scales approximately with $1/\tau$, which can be straight-forwardly shown from Eq.~\ref{eq:MEC_analytical}. Further crucial parameters for the minimum energy curve are the optical near field decay length $\zeta$ and the peak optical field strength (see Extended Data Fig.~\ref{fig:NILES_amplitude} and Methods). While the optical near field decay is the root cause for NILES, strong static electric fields often present at metal needle tips can further influence the precise shape (see Methods).

\section*{Extracting electron momentum widths}
Thus far we discussed NILES based on classical trajectories. In an extended semi-classical model, we can include quantum features, by, e.g., assigning a finite momentum width~\cite{Krger2012_2} to the electron trajectories to account for the finite birth time window within the field cycle~\cite{Kim2023, Dienstbier2023}. To this end, each trajectory contributes to the final spectrum via a Gaussian momentum distribution with a width $\sigma_{p}$ centered at the final momentum of the classical trajectory. We determine the optimal value of this width by fitting the simulated spectra to the experiment (see Methods). In particular, this process allows us to match the sharpness of the measured NILES features. The resulting width of $\sigma_p = \SI{4.8e-26}{\kilo\gram\meter\per\second} = 0.024$\,a.u. is considerably smaller than that predicted by conventional models for atomic ATI spectra $\sigma_{p,\mathrm{theory}} \sim 0.4-0.5$\,a.u.~\cite{Popov1999, Krger2012}. Hence, NILES calls for a description beyond conventional broadening concepts and may enable investigating quantum diffusion effects and momentum broadening at low energies in a so far unexplored scenario, representing an exciting outlook of forthcoming NILES-based experiments.

\section*{Individual attosecond electron burst}
\begin{figure}
\includegraphics[width=0.99\linewidth]{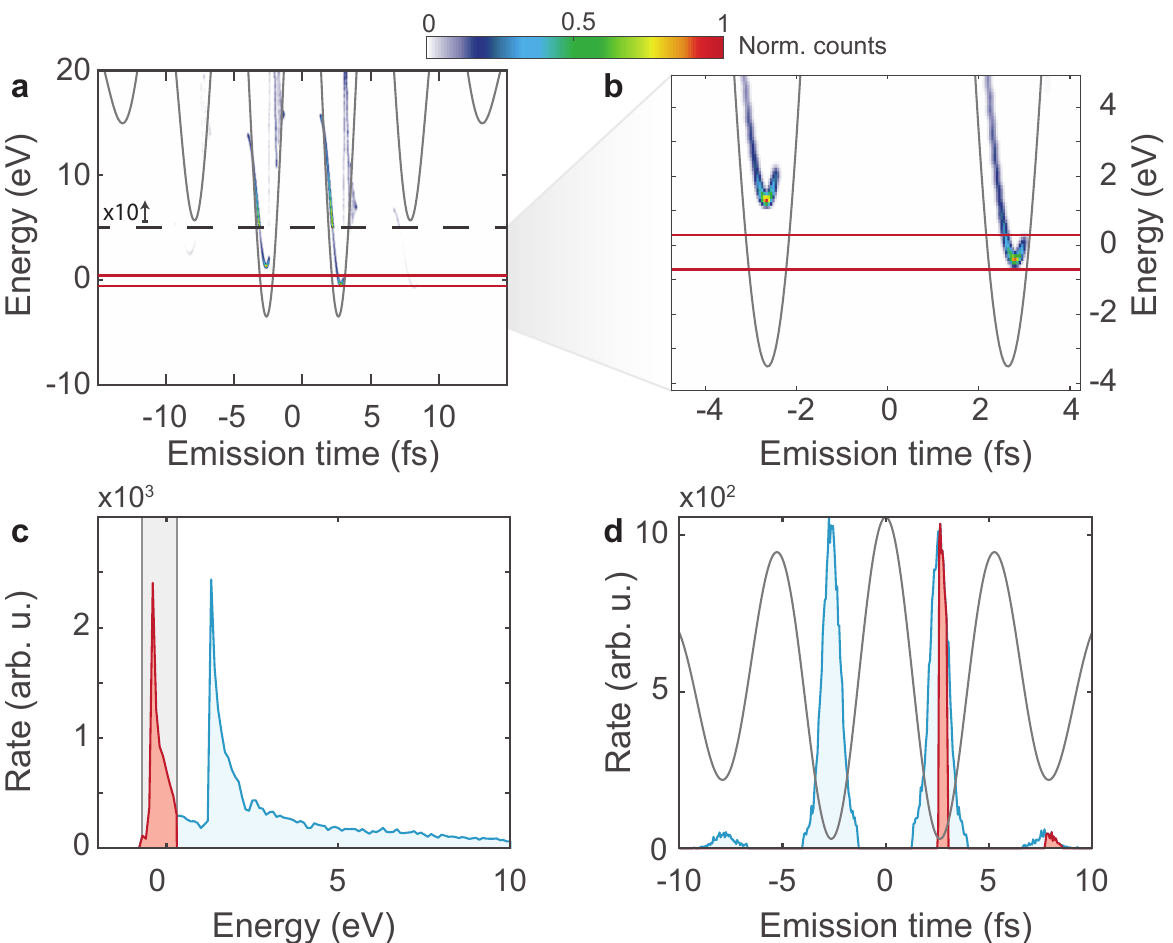}
\caption{\textbf{Isolation of individual attosecond electron bursts through NILES filtering}: \textbf{a} Energy of the emitted electrons as function of the emission time for parameters similar to Fig.~\ref{fig:niles-scheme}c, obtained from the three step model. The color code indicates the emission yield, obtained from the Yudin-Ivanov rate \cite{Yudin2001}. We note that the values above 5\,eV were multiplied by a factor of 10 for better visibility. \textbf{b} The zoom-in of \textbf{a} shows that the cycle-dependent energy shift through NILES allows us to single out electron emission from one cycle when filtering an energy band from -0.3 to 0.7\,eV (red lines). \textbf{c} Resulting electron energy spectra. The red area again shows the filtered electrons. \textbf{d} Emission time of the energy-filtered electrons (red) and all electrons (blue). With these filter settings 17\,\% of the emitted electrons are confined within a window of $430$\,as. \SI{6}{\percent} of the filtered electrons are in another cycle. The simulations include the bias voltage applied to the tip, but no spectral blurring filter as in the simulation data shown Fig.~\ref{fig:niles-scheme} \label{fig:as_pulse}}
\end{figure}

As shown by simulation results in Fig.~\ref{fig:explanation_NILES_Rostock}c,d, NILES leads to an emission time-sensitive energy shift of the direct electrons within the laser {\it pulse}. This shift allows us to select electrons from a specific cycle just by filtering the associated energies, shown exemplarily as red band in Fig.~\ref{fig:as_pulse}a,b. For this energy band, ranging from -0.7 to 0.3\,eV, other dominant emission cycles are energetically well separated. This selection of the energy (Fig.~\ref{fig:as_pulse}c) thus translates to an emission time filtering. Fig.~\ref{fig:as_pulse}d shows the emission times of the filtered electrons (red) versus all electrons (blue). Clearly, from the multi-cycle laser pulses, we isolate an individual \SI{430}{\atto\second} long electron emission burst in the simulation. This burst contains a substantial \SI{17}{\percent} of the emitted electrons. Energy filtering is readily available in electron microscopes, facilitating to single out an individual attosecond electron emission burst. To harness this temporal confinement, available technologies for electron dispersion compensation have to be employed~\cite{Gliserin2015,Priebe2017,Schoenenberger2019,Black2019}. We note that the exact width of the emission window depends strongly on the experimental conditions, the filtering and the intrinsic energy width of the electron emitter. Last (see Methods), we note that we expect this work to result in integrated or even on-chip CEP measurement and stabilisation devices because of the large CEP-dependent current.

\FloatBarrier
\appendix

\newpage
\bibliography{A3_literature}

\FloatBarrier

\setcounter{figure}{0} 
\renewcommand{\figurename}{Extended Data Fig.}
\include{A2_SI2}

\subsection*{Acknowledgments} 
We acknowledge Philip Dienstbier for discussions. This research was supported by the European Research Council (Advanced Grant AccelOnChip), the Deutsche Forschungsgemeinschaft (DFG, German Research Foundation) through TRR 306 QuCoLiMa ("Quantum Cooperativity of Light and Matter"), CRC 953 ("Synthetic Carbon Allotropes"), CRC 1477 ("Light-Matter Interactions at Interfaces", grant no. ID 441234705), and the FET Open project PetaCOM of the EU. J.H.\ acknowledges funding from the Max Planck School of Photonics.

\subsection*{Author contributions}
S.M.\ and J.H.\ measured and analyzed the data and performed simulations. F.L.H., D.L.,  S.W.\ and T.W.\ set up the laser system. F.L.H.\ assisted in data taking. A.H., L.S.\ and T.F. did the analysis of full and drift trajectories and of the momentum width. S.H.\ fabricated the FIB-milled tip. S.M., J.H.\ and P.H.\ wrote the manuscript with input from all authors. 


\end{document}

%% file: A2_SI2.tex
\section*{Methods}\label{secA1}
\subsection*{Experiment}

\begin{figure*}[h]
\includegraphics[width=\linewidth]{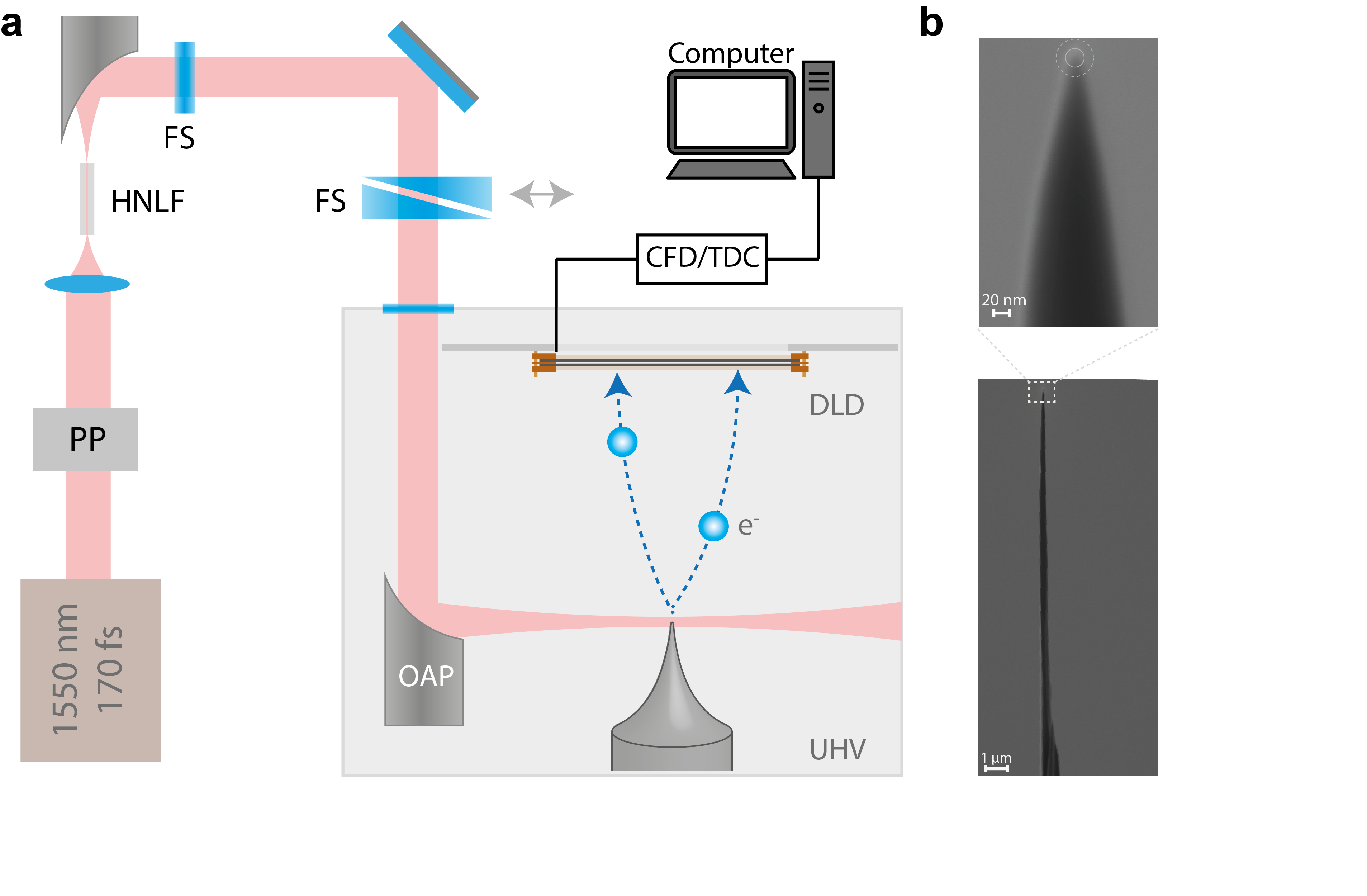}
\caption{\textbf{Experimental setup.} \textbf{a} Laser and ultra-high vacuum (UHV) system. PP: Pulse picker based on a Pockels cell. HNLF: highly non-linear fibre. FS: fused silica glass for dispersion compensation.  OAP: off-axis parabolic mirror, mounted inside of the UHV chamber. DLD: delay-line detector with micro-channel plate in front, allowing single electron detection. CFD: constant fraction discriminator. TDC: time-to-digital converter. Electrons (blue) laser-emitted from the tungsten needle tip are accelerated to the electron detector. \textbf{b} Scanning electron micrograph of tip after FIB milling with close-up of the tip apex. {\color{black_red}The inner solid circle corresponds to a radius of curvature of $R =\SI{10}{\nano\meter}$ and the outer one to $R = \SI{20}{\nano\meter}$}. For more details see text.}
\label{fig:experimental_setup}
\end{figure*}

We use an amplified erbium fiber oscillator system as laser source with 170\,fs pulse duration operating at a repetition rate of 80\,MHz. We reduce the repetition rate with a pulse picker (PP) down to 100\,kHz (Extended Data Fig.~\ref{fig:experimental_setup}). We broaden the spectrum by a highly nonlinear normal dispersion fiber (HNLF)\cite{Lesko2021}. After compression with fused silica (FS) we obtain a pulse duration of 11.5\,fs measured by frequency-resolved optical gating (FROG). The laser pulses are passively CEP-stable, and we vary the CEP by two motorized glass wedges in the beam path.

These laser pulses are then focused by an off-axis parabolic mirror ($f=\SI{15}{\milli\meter}$) to a focal spot size of \SI{3}{\micro\meter} ($1/e^2$-intensity radius) onto a tungsten needle tip. The tip is placed inside an ultrahigh vacuum chamber at a pressure of \SI{8e-10}{\hecto\pascal}. For the measurement shown in Fig.~\ref{fig:experiment_NILES}, we use an electro-chemically etched tungsten tip post-processed by focused ion beam milling (FIB), leading to even cleaner electron spectra. The NILES feature also shows up in just-etched tungsten tips; see Extended Data Fig. \ref{fig:absolute_phase}, which is recorded with a non-FIBed tip. To remove remnants of oxides at the very tip apex, we used in-situ cleaning by field evaporation before the experiments in all cases~\cite{Mller1955}.

The laser-triggered electrons are accelerated from the negatively biased tip ($U_\mathrm{tip} = \SI{-30}{\volt}$) towards a delay-line detector (DLD). By the time-of-flight and the position of the electrons at the detector we calculate the energy of each individual electron and record electron energy spectra. Typically we choose the laser power such that the electron count rate lies around 0.1 to 0.5 electrons per pulse on average. In this way most of the electron events are single electrons and potential Coulomb effects are suppressed \cite{Meier2023}.

For the measurement in Fig.~\ref{fig:experiment_NILES} we used a step size of \SI{0.26}{\radian} in CEP and recorded $10^5$ electrons per phase step over a range of $\sim$$1.6 \times 2 \pi$ at a near field peak intensity of \SI{2.1e13}{\watt\per\square\centi\meter}. We determine this intensity from the measured cut-off position of \SI{42}{\electronvolt}. 

{\color{black_red}Depending on the exact count rate and the CEP range we can record a full map like in Fig.~\ref{fig:experiment_NILES} on the few-minute time scale.  We note, however, that the electron emission yield and the spectra are typically very stable on the hour time scale. Here we profit from the low repetition rate of our laser system compared to typical oscillators operating around \SI{80}{\mega\hertz}, which leads to less heating effects and consequently longer lifetimes of the metal needle tips.}

We match the CEP axes (offset) of experiment and theory based on a two-dimensional convolution of the low-energy region, exactly where the NILES feature shows up, similar to schemes that employ the cut-off position or the photo-ionization yield \cite{Haworth2006,Kling2008, Lan2008,Wittmann2009,Krger2011,Sayler2015}. The resulting phase error of this scheme can be as low as \SI{120}{\milli\radian}.

\subsection*{Analytical Form of the Minimum Energy Curve}
{\color{black}
We consider the ponderomotive energy at the tip surface (i.e. in the maximally enhanced field and defined for peak field strength)

\begin{linenomath}
\begin{align*}
U_\mathrm{P} = \xi_0^2 U_\mathrm{P}^{\mathrm{inc}} = \xi_0^2 \frac{e^2E_0^2}{4m\omega^2} = \xi_0^2 \frac{e^2 I_0}{2c\epsilon_0 m \omega^2},
\end{align*}
\end{linenomath}
where $U_\mathrm{P}^{\mathrm{inc}}$ is the ponderomotive energy of the incident laser pulse without field enhancement.

Considering the temporal intensity envelope and the spatial near-field profile, the dynamical (spatio-temporal) ponderomotive energy reads

\begin{linenomath}
\begin{align*}
U_\mathrm{P}(x,t) = \xi(x)^2 \frac{e^2I(t)}{2c\epsilon_0m\omega^2}.
\end{align*}
\end{linenomath}

This results in the (also spatio-temporal) ponderomotive force

\begin{linenomath}
\begin{align*}
F_\mathrm{P}(x,t) = -\diff{}{x} U_\mathrm{P}(x,t) = -\frac{e^2I(t)}{2c\epsilon_0 m \omega^2}\cdot 2 \xi(x) \xi'(x).
\end{align*} 
\end{linenomath}
Solving this equation yields the drift trajectories shown in Fig.~\ref{fig:explanation_NILES_Rostock}, which include the full near field decay but neglect the oscillations of the optical field because we average over one optical cycle. Note that electrons start one quiver length away from the surface in the drift picture.

As an upper estimate, we now only evaluate the ponderomotive force at the surface, where $\xi(x=0) = \xi_0$ and $\xi'(x=0) = \xi'_0$, and find the temporal ponderomotive force at the surface 
\begin{linenomath}
\begin{align*}
F_\mathrm{P}^\mathrm{surf}(t) = -\frac{e^2I(t)}{2c\epsilon_0m\omega^2}\cdot 2 \xi_0 \xi_0'.
\end{align*}
\end{linenomath}

The final momentum of an electron starting at rest at birth time $t_b$ follows from the integration of the classical equation of motion and yields
\begin{linenomath}
\begin{align*}
p_\mathrm{final}(t_b) = -2 U_\mathrm{P}^\mathrm{inc} \xi_0\xi_0'\mathcal{F}(t_b),
\end{align*}
\end{linenomath}
where $\mathcal{F}(t_b) = \frac{1}{I_0}\int_{t_b}^\infty I(t) \mathrm{d}t$ is the remaining normalized pulse fluence the electron may experience after its birth into the optical field. 

\noindent The drift energy of the electrons then reads 
\begin{linenomath}
\begin{align*}
E_\mathrm{final}(t_b) = \frac{2}{m}\left(U_\mathrm{P}^\mathrm{inc}\xi_0\xi_0'\right)^2 \mathcal{F}^2(t_b).
\end{align*}
\end{linenomath}
}

\subsection*{Three step model simulations}
One common way to simulate the electron energies of electrons triggered from metal needle tips is based on the three-step model  \cite{Corkum1993,Schafer1993,Krger2012_2}. It is based on a {\color{black_red} 3D} semi-classical point-particle trajectory analysis and allows us to simulate the electron energies for different laser intensities, as function of carrier-envelope phase, and for spatially inhomogeneous optical fields. The initial conditions of the electrons in the three spatial dimensions are given by the projection of a two-dimensional Gaussian distribution on a hemisphere to model the tip. The starting times and the yield of the electrons are based on a closed-form quantum-mechanical model, {\color{black_red} often called Yudin-Ivanov (YI) rate} \cite{Yudin2001, Shi2021}. {\color{black_red} Originally this model was developed for the electron emission from atoms. The metal-vacuum interface at tips breaks the symmetry as compared to an atom. We assume, therefore, that electron emission takes only place in negative half-cycles as we are in the sub-cycle tunneling regime (see also Fig.~\ref{fig:as_pulse}d). To avoid Coulomb repulsion effects, we artificially set the total rate to one electron per pulse in these simulations}. After the emission the electrons are propagated classically by numerically integrating the equations of motion {\color{black_red}based on a standard fourth-order Runge-Kutta scheme}. Here we take into account both the static field applied to the tip as well as the oscillating field.
We model the static field by a spherical capacitor with the outer sphere set to infinity \cite{Tsujino2018}. We choose \(U_\mathrm{tip}=\SI{-10}{\volt}\) as bias voltage for {\color{black_red}all} our simulations {\color{black_red} if not specified otherwise}. This voltage leads to a static field of \SI{1}{\volt\per\nano\meter} at the apex of the tip.
For the oscillating near field, we choose a pulse with Gaussian envelope representing the laser pulse given by

\begin{linenomath}
\begin{align*}
    E(r,t) ={}& \exp{\left(-\frac{2\ln(2)t^2}{\tau^2}\right)} \cos{\left(\omega t + \Phi_\mathrm{CEP}\right)} \\
    &\cdot \left[1 + (\xi_0 - 1) \exp{\left(-\frac{r}{\zeta}\right)}\right].
\end{align*}
\end{linenomath}
Here, the pulse duration \(\tau\) is given as the intensity full width at half maximum (FWHM). We chose a near field shape with a decay constant of $\zeta=$\SI{10}{\nano\meter} unless stated otherwise. For the simulations shown in Fig.~\ref{fig:niles-scheme}a-c, the field enhancement factors chosen for the simulation are $\xi_0 = 1$ (Fig.~\ref{fig:niles-scheme}a), $\xi_0 = 7$ (Fig.~\ref{fig:niles-scheme}b) and $\xi_0 = 15$ (Fig.~\ref{fig:niles-scheme}c). We chose these values to account for the fact that the field enhancement factor increases for decreasing tip radius.

\begin{figure*}[h]
\includegraphics[width=\linewidth]{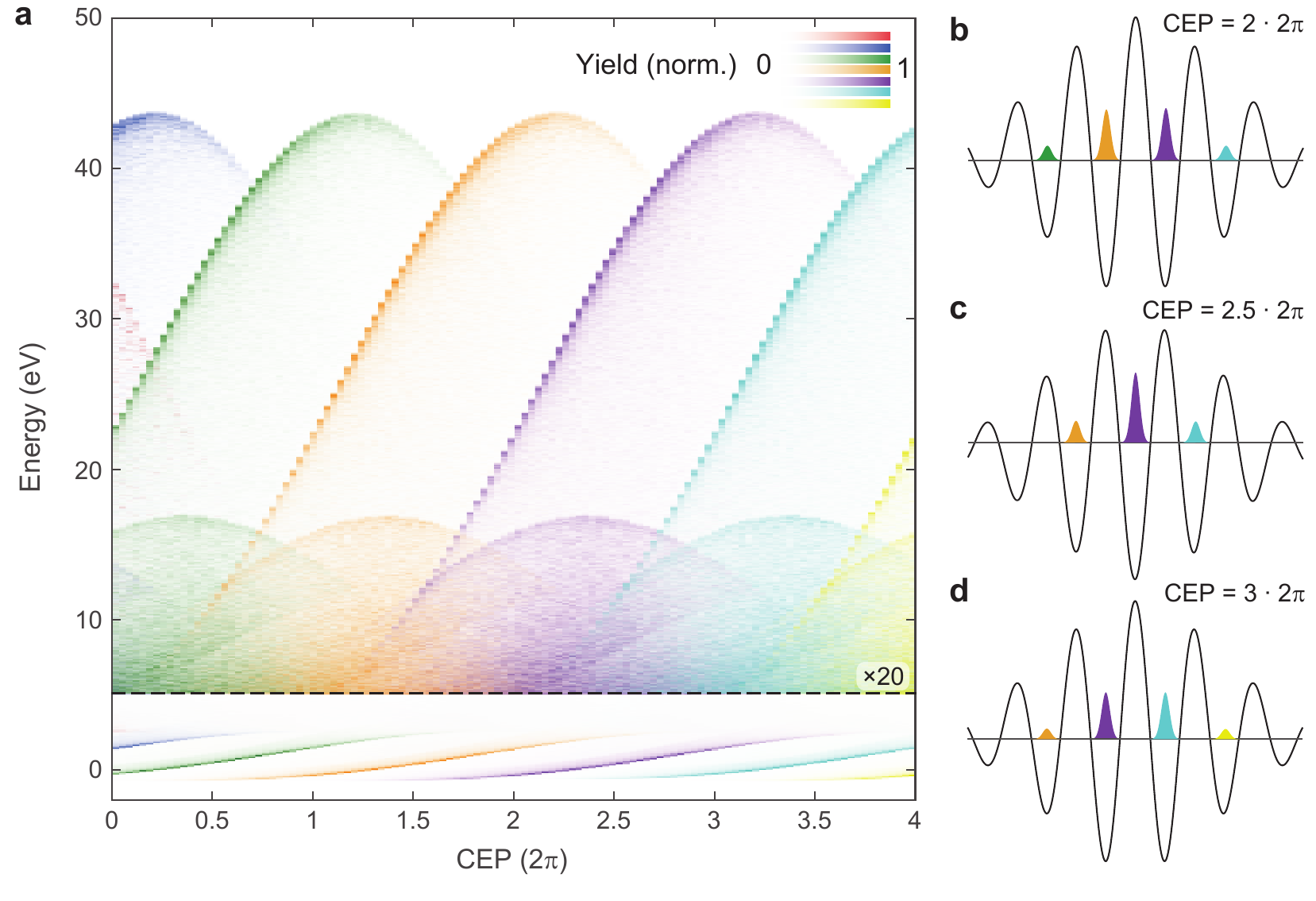}
\caption{\textbf{The contribution of different emission cycles to the full spectrum as a function of the CEP.} \textbf{a} The kinetic energy as a function of CEP is shown in color according to the respective emission cycles shown in b-d. For better visibility, all values above \SI{5}{\electronvolt} have been multiplied by a factor of 20. \textbf{b} At a CEP of $2\cdot(2\pi)$, four cycles contribute to the electron emission in \textbf{a}, shown schematically, with the two outer cycles (green, teal) having only a small contribution compared to the two centered emission bursts (orange, purple). \textbf{c-d} The emission bursts are shifted to earlier times within the laser pulse as the CEP is increased.}
\label{fig:abstract_cycles}
\end{figure*}

In Extended Data Fig.~\ref{fig:abstract_cycles}a we show the CEP-resolved energy spectrum as obtained from the simulation. We color-code the emission distributions from the different optical cycles, where each color corresponds to a specific cycle and the rate is given by the opacity of each color. In Extended Data Fig.~\ref{fig:abstract_cycles}b-d we depict which optical cycle corresponds to which color for three different CEP values. This illustration clarifies that the number of NILES corresponds to the number of emitting cycles.
Furthermore, we find that the lowest energy part from $-1$ to $1$\,eV is dominated by the purple emission burst for a CEP of $\sim$$1.7\cdot 2\pi$ to $2.3\cdot 2\pi$, for example. This is the basis to obtain a single attosecond emission burst, as shown in Fig.~\ref{fig:as_pulse}.\\


\subsection*{Parameter dependencies of the minimum energy curve}
In Fig.~\ref{fig:explanation_NILES_Rostock}e we show how NILES arise for one set of parameters{\color{black_red}, when the near field is inhomogeneous}. There, the minimum energy curve (solid lines), representing the origin of NILES, is influenced by all parameters that change the optical field the electron experiences while it performs a quiver motion. 

\begin{figure*}[h]
\includegraphics[width=0.9\linewidth]{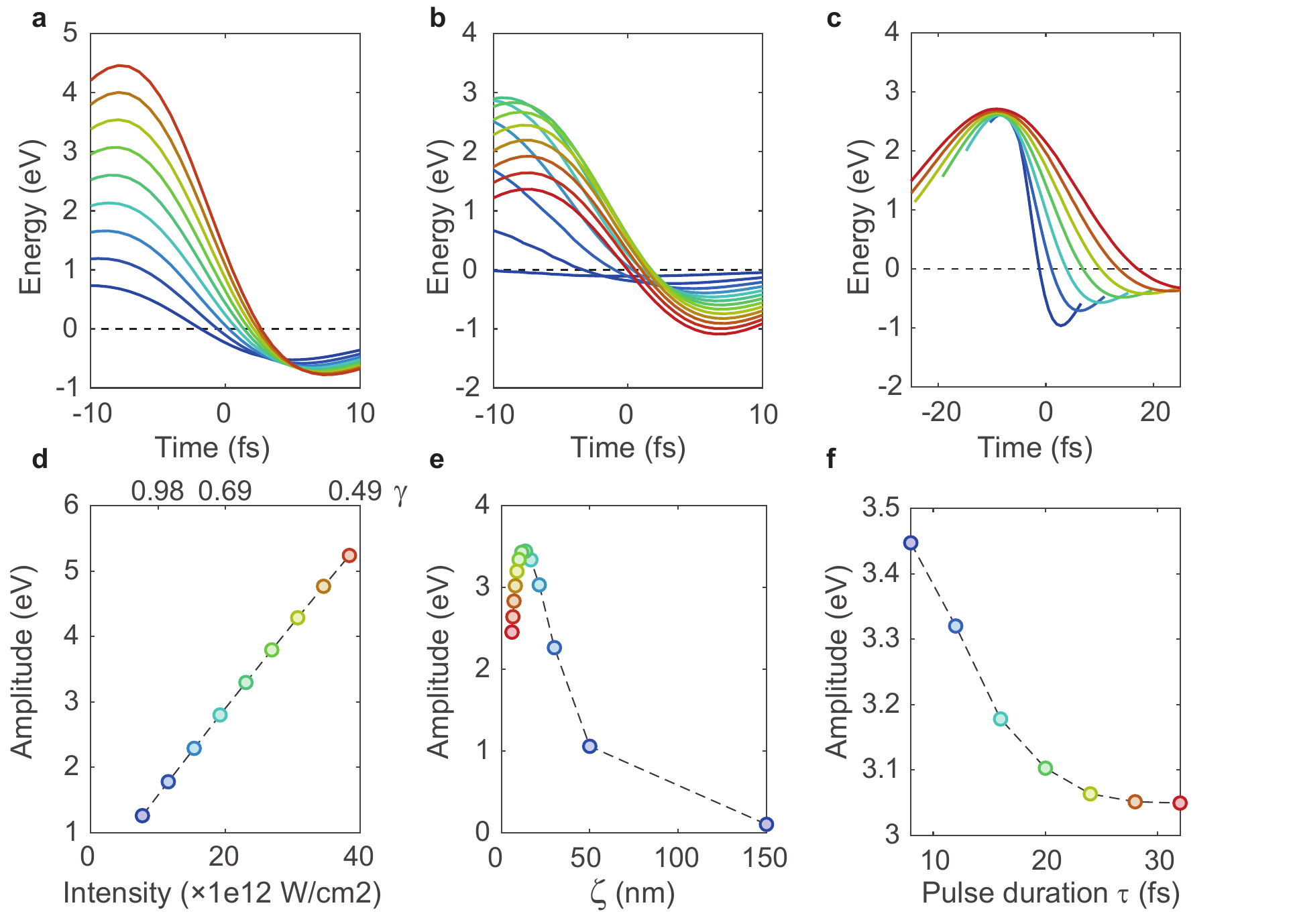}
\caption{\textbf{Minimum energy curve for various parameter sets}. Simulated minimum energy curves based on the semi-classical simulation for different \textbf{a} intensities,  \textbf{b} near field decay lengths $\zeta$ and \textbf{c} pulse durations $\tau$. The color scale is matched in each plot \textbf{a-c} to the one below \textbf{d-f}, where the value for each color can be read on the horizontal axis. \textbf{d-f} show the  amplitude of the minimum energy curve, defined as the minimum energy subtracted from the maximum energy for each line in \textbf{a-c}. {\color{black_red} In \textbf{d} we additionally show the Keldysh parameter $\gamma$ in a second abscissa at the top.} See text for details.}
\label{fig:NILES_amplitude}
\end{figure*}

In Extended Data Fig.~\ref{fig:NILES_amplitude}a-c we show the dependence of the minimum energy curve on the most relevant parameters: peak optical near-field intensity, near field decay length $\zeta$ and laser pulse duration $\tau$. We discuss these parameters in the following. We use a near field decay length of $\zeta =\SI{10}{\nano\meter}$, a pulse duration of \(\tau=\SI{11.5}{\femto\second}\), a near-field intensity of \SI{2.3e13}{\watt\per\square\centi\meter}, a central wavelength of \SI{1570}{\nano\meter} and a tip voltage of \SI{-10}{\volt}, if not stated otherwise.

In order to quantify the effect of these parameters on NILES, we further calculate the maximum amplitude for each minimum energy curve, given by the difference of maximum energy and minimum energy. We refer to this amplitude as the {\it minimum energy curve amplitude} and show these amplitudes in Extended Data Fig.~\ref{fig:NILES_amplitude}d-f.

We find that the minimum energy amplitude increases for increasing intensity (Extended Data Fig.~\ref{fig:NILES_amplitude}a). Extended Data Fig.~\ref{fig:NILES_amplitude}d shows that it does so in a linear fashion. This increase reflects that the electrons are driven further away from the surface in the optical near-field with increasing intensity, thus they experience a larger variation of the optical near-field leading to an enhanced NILES effect. We note that this linear scaling is similar to the scaling of the low-energy peak for one fixed emission time, investigated by Sch\"otz et al. \cite{Schtz2018}. {\color{black_red} We further note that for our set of parameter NILES is not detectable in the multiphoton region, as the amplitude would be too small to be detected for \(\gamma>2\).}

When we sweep the near field decay length $\zeta$, from $\zeta = \SI{150}{\nano\meter}$ (dark blue) towards smaller decay lengths (Extended Data Fig.~\ref{fig:NILES_amplitude}b), we observe first an increase in the minimum energy curve amplitude, down to a value of $\zeta=\SI{13.3}{\nano\meter}$ (green curve). When $\zeta$ is further decreased, however, the amplitude reduces again. Physically, this reduction is a consequence of an increasingly strong quenching of the quiver motion for smaller decay lengths, i.e. the amplitude of the oscillatory motion decreases.

In Extended Data Fig.~\ref{fig:NILES_amplitude}c we show the minimum energy curve for different pulse durations $\tau$. Whereas the amplitude of the minimum energy curve only slightly increases with decreasing pulse duration (Extended Data Fig.~\ref{fig:NILES_amplitude}f), the width of the minimum energy curve clearly broadens for longer pulses. This broadening along the time axis is to be expected, as the minimum energy curve is mainly affected by the time that the electron needs to escape the optical near field, as well as the optical pulse duration. In the limit of an infinitely long pulse, i.e. the continuous wave case, the minimum energy curve would be constant above zero for our case, because without temporal envelope all electrons born with a time difference of one optical period traverse the same near field. In the other case of constant pulse duration and increasing $\zeta$, the NILES amplitude vanishes to zero (Extended Data Fig.~\ref{fig:NILES_amplitude}b).

\subsection*{Influence of static electric fields}
\begin{figure*}[h]
\includegraphics[width=0.7\linewidth]{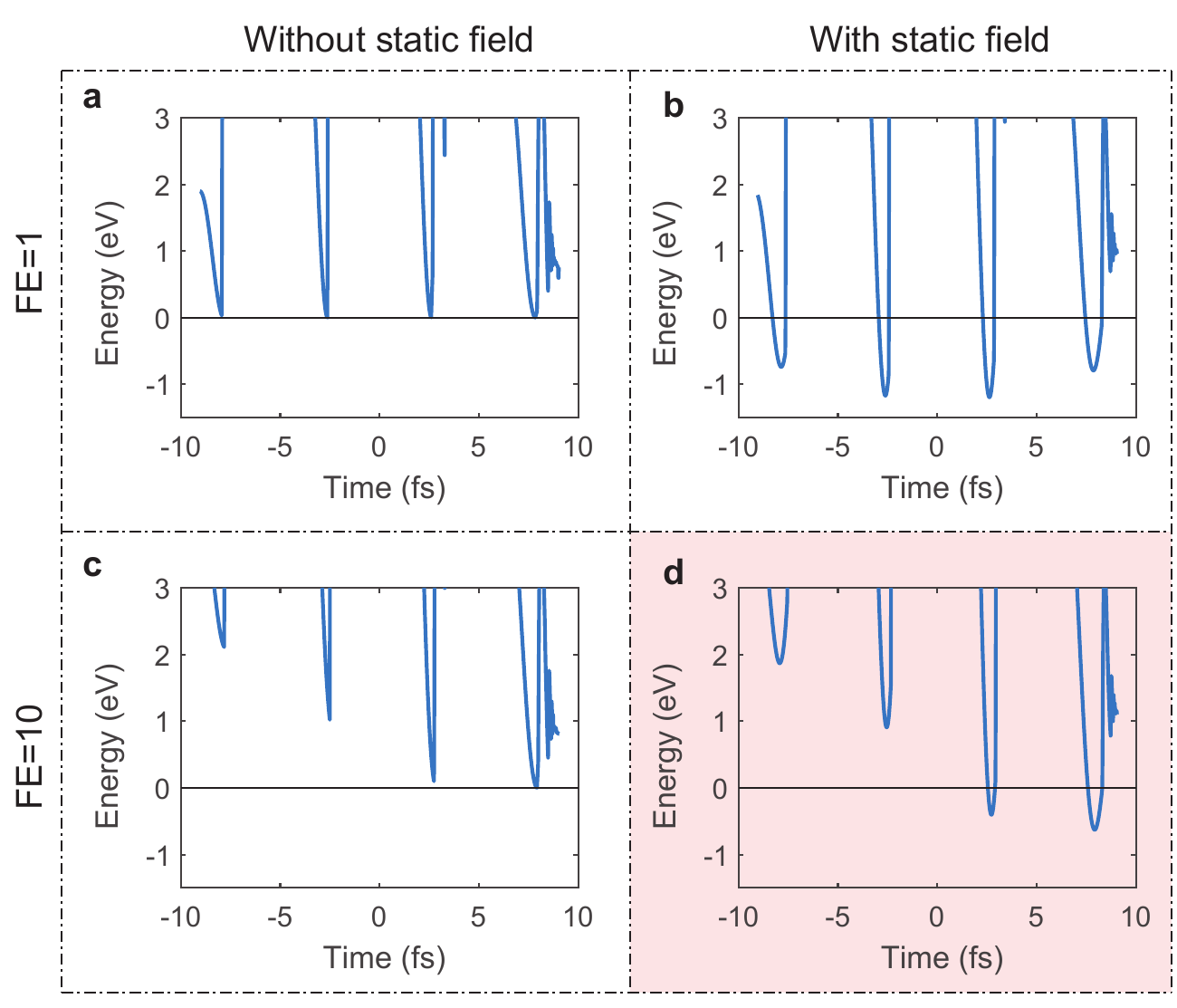}
\caption{\textbf{Influence of the static bias voltage}. Simulations of the final energies of electrons for four cases without (\textbf{a} and \textbf{c}) and with (\textbf{b} and \textbf{d}) a static bias voltage of \SI{10}{\volt} (peak DC near field of \SI{1}{\volt\per\nano\meter}) and with a field enhancement factor of \(\xi_0=1\) (\textbf{a} and \textbf{b}), thus with a homogeneous optical  field ($\zeta = \infty$), and with \(\xi_0=10\) and $\zeta = \SI{10}{\nano\meter}$ (\textbf{c} and \textbf{d}). We plot only the low-energy part from \SI{-1.5}{\electronvolt} to \SI{3}{\electronvolt}. For each simulation, we chose a CEP of zero and starting times of \SI{-10}{\femto\second} to \SI{10}{\femto\second}. The case for a typical needle tip present in our experiments is highlighted with a red background (\textbf{d}). The differences of the fours panels show clearly that the presence of both optical {\it and} DC near field are required to fully understand the electron dynamics.}
\label{fig:DC_field_effect}
\end{figure*}

NILES is an effect entirely based on the quickly decaying optical near field. However, the shape and the amplitude of NILES are influenced by the static bias field at the tip apex occurring when biasing the tip with a voltage $U_\mathrm{tip}$. This voltage results in an additional contribution to the final electron energies of $E_\mathrm{DC} = e\cdot U_\mathrm{tip}$ which we subtract when displaying the energies in our plots. To show this influence in more detail, we simulate the final energies of electron trajectories for four different cases: with and without a static field for a homogeneous field ($\zeta=\infty$, $\xi_0=1$) and a decaying near field ($\zeta=\SI{10}{\nano\meter}$, $\xi_0=10$) . The low-energy part of these simulations, shown in Extended Data Fig.~\ref{fig:DC_field_effect}, reveals that only in the presence of a the static field, energies below zero (i.e. final kinetic energies below $E_\mathrm{DC}$) can be reached.

This result may come surprising when one thinks about an electron emitted without a laser field that surfs down only the static potential, leading to an energy of $E_\mathrm{DC}$ (in our case \SI{10}{\electronvolt}). It is the interplay of the laser induced near-field and the static field that enables emission with final energies below the DC voltage.
However, this effect can only be observed clearly in the case of a strong inhomogeneity of the static field, given by the small dimensions of the tip.

\begin{figure*}[ht!]
\includegraphics[width=1\linewidth]{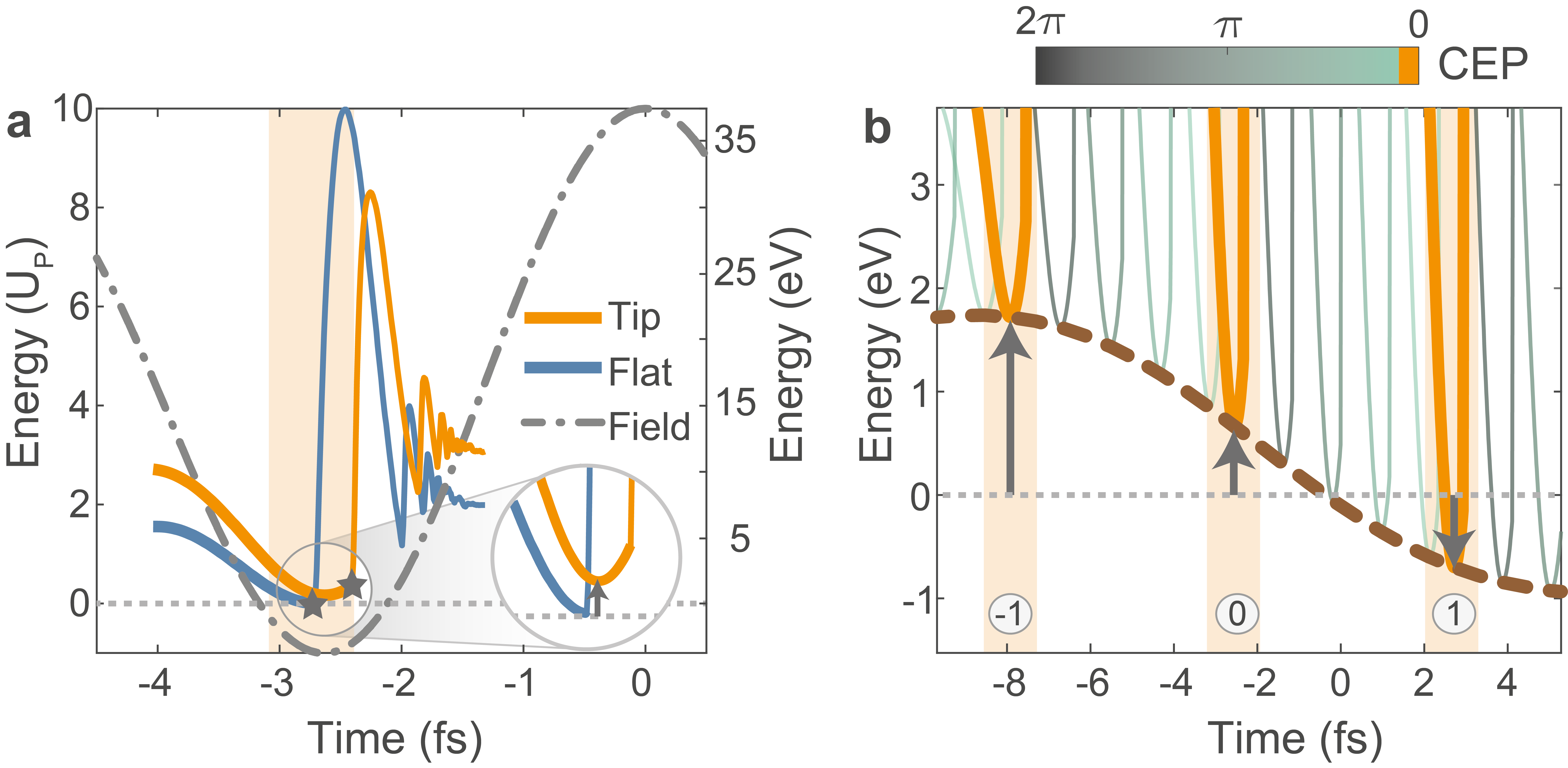}
\caption{
\textbf{Shape of NILES}:
 \textbf{a} Final energies of electrons as function of their emission time for one cycle extracted from the semi-classical three step model, corresponding to case 2 in Fig.{\color{black_red}~\ref{fig:niles-scheme}}a. The stars mark where the transition from direct (bold) to rescattering (thin) trajectories happens (left of the star are the direct electrons). The electron near field motion leads to a non-zero minimal energy, whereas without static and optical near field the minimal energy is always 0. \textbf{b} Bottom parts of curves like shown in \textbf{a} for various CEPs and covering emission times in multiple cycles. Clearly, the minimum energies follow a smooth curve, the minimum energy curve, given by the brown dashed curve. It is directly reflected in the experimentally observable NILES. In this Figure, the color scale for the CEP is $2\pi$-periodic because the the emission time axis makes the cycles unique.  
  \label{fig:shape_of_NILES}}
\end{figure*}

{\color{black_red} We further show in analogy to the minimum energy curve in Fig.~\ref{fig:explanation_NILES_Rostock}e, the result when the static potential like in the experiment is included (Extended Data Fig.~\ref{fig:shape_of_NILES}).}

{\color{black}
\subsection*{Extraction of momentum width}

}

\begin{figure*}[h]
\includegraphics[width=0.8\linewidth]{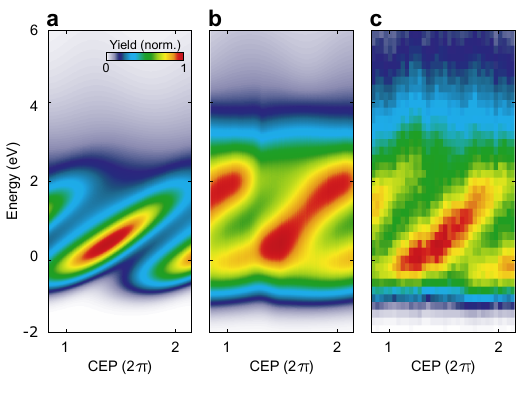}
\caption{{\color{black_red}\textbf{Momentum width extraction from NILES.} \textbf{a} Simulated low energy region on a linear scale, similar to Fig.~\ref{fig:niles-scheme}. \textbf{b} Same region simulated but now all calculated trajectories are associated with a momentum width of $\sigma_p = \SI{4.8e-26}{\kilo\gram\meter\per\second} = 0.024$\,a.u. This value is found from optimization algorithms detailed in the text. \textbf{c} Measured NILES for comparison, same as in Fig.~\ref{fig:experiment_NILES}. Clearly, the inclusion of a momentum width leads to a high agreement between experiment and simulation.}}
\label{fig:extraction_momentum_width}
\end{figure*}

{\color{black}For the extraction of the momentum width we simulated spectra similar as detailed above with minor modifications to the model. First, we weighted trajectories using the WKB-tunneling rate. Second, we include the experimental energy resolution of the detector in the simulations. The resulting low energy spectral features (cf. Extended Data Fig.~\ref{fig:extraction_momentum_width}a) qualitatively capture the experimental NILES signatures (Extended Data Fig.~\ref{fig:extraction_momentum_width}c) but do not agree quantitatively. For a quantitative reconstruction, we further modify our simulations and perform an optimization (Nelder-Mead simplex method) for three additional parameters: First, the width $\sigma_p$ of a Gaussian momentum distribution that we associate with each trajectory \textit{after} propagation. Second, the exponent $n$ of a now considered exponential ionization rate $R = E^n(x=0,t)$, where $E(x=0,t)$ is the locally enhanced electric field at the surface. Third, a random CEP-shift $\varphi_\text{CE}$. In this way, we optimized for the best overlap between the simulated and measured low energy region of the spectra, including the NILES features. Out of 100 initial conditions for the optimization the best agreement (compare Figs.~\ref{fig:extraction_momentum_width}b,c) was found for $\sigma_p = \SI{4.8e-26}{\kilo\gram\meter\per\second} = 0.024$\,a.u., $n = 4.24$ and $\varphi_\textit{CE} = 0.05\pi$. Especially including the momentum width paves the way to examine inter- and intra-cycle interference effects to a large extent with extended semi-classical simulations, for example \cite{Krger2012_2}.}


\subsection*{CEP-dependent current}
A CEP-dependent current is of highest interest for direct CEP-locking of lasers without the need of non-linear interferometers. While CEP-dependent currents have been demonstrated, so far they are too small for direct CEP locking \cite{Krger2011, Piglosiewicz2013,Rybka2016,Keathley2019}. In the following we illustrate how NILES can be used to realize such a device, based on the {\it direct} electrons and thus with a much larger fraction of electrons contributing to the CEP-dependent current. 

\begin{figure*}[h]
\includegraphics[width=\linewidth]{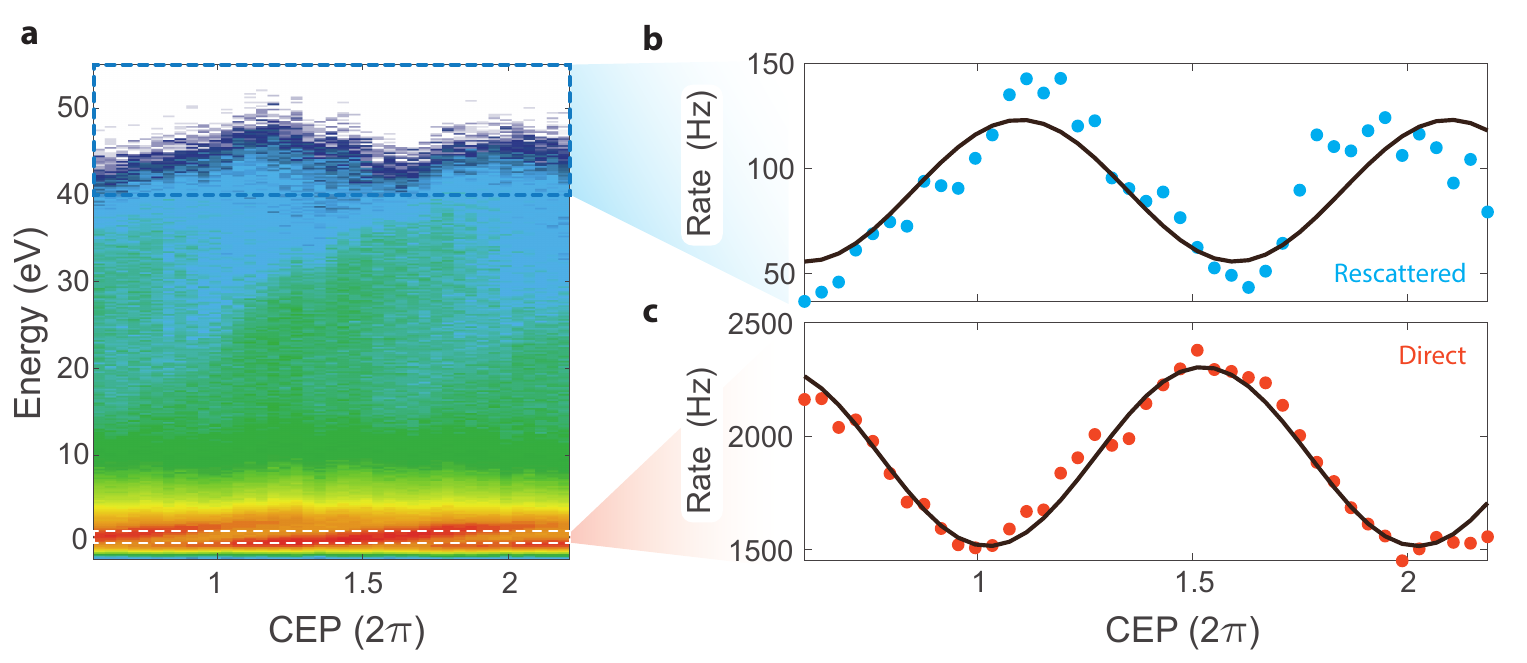}
\caption{\textbf{Direct electron CEP sensor.} \textbf{a} CEP-resolved energy spectra from Fig.~\ref{fig:experiment_NILES}. \textbf{b} Current as function of CEP obtained by energy filtering: rescattered electrons (blue) and direct electrons (red); for the rescattered electrons we chose $40-60$\,eV and for the direct electrons $0.4-1.2$\,eV, indicated by dashed boxes in \textbf{a}. The rescattered electrons show a higher modulation depth of 46\,\% versus 33\,\% for the direct electrons, extracted by a sine-fit (black). However, the yield of the direct electrons is higher by a factor 17, resulting in a better signal to noise ratio. See text for details. 
}
\label{fig:CEP_current}
\end{figure*}

Extended Data Fig.~\ref{fig:CEP_current}a shows again the results of the measurement similar to Fig.~\ref{fig:experiment_NILES}b, from which we select two energy bands for the rescattered (blue) and the direct electrons (red). The chosen energy band for the rescattered electrons is 40-60\,eV and 0.4-1.2\,eV for the direct electrons.
In Extended Data Fig.~\ref{fig:CEP_current}b,c we show the integrated yield of the electrons of these bands, which both show a sine modulation as function of the CEP. Whereas the modulation depth for the direct electrons is 33\% and thus smaller than the 46\% for the rescattered electrons, the total yield of the direct electrons is 17 times higher than that of the rescattered electrons. Hence, in total NILES yields a $\sim$9 times better signal-to-noise ratio (SNR) than the rescattered electrons.

To stabilize a laser oscillator a SNR of 30\,dB at a bandwidth of 100\,kHz is required \cite{Borchers2014,Dienstbier2019}. Assuming this bandwidth and a repetition of 80\,MHz, we obtain a SNR of 8\,dB for NILES, too low for a stabilization by factor of over 100 in current. However, using a nanotip array the current can be directly enhanced by orders of magnitude \cite{Swanwick2014}. We foresee that this scheme can even be miniaturized to fit into the package of a small standard photodiode. {\color{black_red} There are already systems that measure the CEP based on the electron emission in planar tip or bow-tie antenna configurations \cite{Keathley2019,Yang2020,ritzkowsky2023large}. Especially for more than two-cycle laser pulses the \textit{integrated} CEP-modulated current of such devices is small, leading to a small SNR. By using NILES and energy filtering, the sensitivity of such schemes can be considerably improved. Choosing an ever larger energy window for the low energy electrons in Extended Data Fig.~\ref{fig:CEP_current}, for example, the CEP-modulated current decreases - which is exactly what happens when measuring the integrated CEP-current.
Further, it was shown that for certain peak intensities the CEP-sensitivity of the integrated current can drop by over an order of magnitude, so-called \textit{vanishing points} \cite{Keathley2019}. This drop results from equally strongly emitting cycles within the laser pulse for different CEPs. Because of the energy shift of different cycles, NILES can energetically separate different cycles and help to avoid such CEP-insensitive points.}

\subsection*{The effect of the near field on the \textit{rescattered} electrons}

\begin{figure*}[ht]
\includegraphics[width=0.95\linewidth]{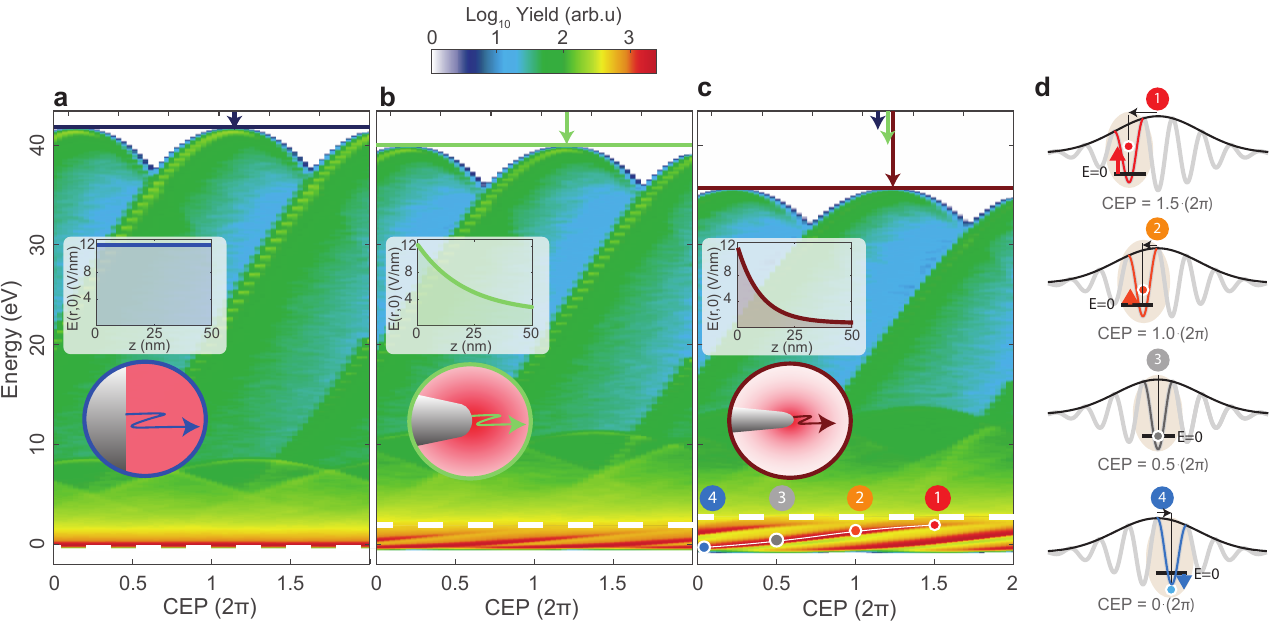}
\caption{\textbf{NILES for three different optical near field decay lengths}: \textbf{a}-\textbf{c} The same semi-classical simulation of electron spectra for three tip radii, with fixed enhanced peak intensities as in Fig.~\ref{fig:niles-scheme}. As before NILES only shows up for the case of inhomogeneous optical near fields. Additionally, we observe that the optical near fields lead to a reduction of the highest energies of the plateau electrons and at which CEP they appear, indicated by the arrows and solid lines. Once more, we stress the excellent agreement of the numerical results in (c) and the experimental data in Fig.~\ref{fig:experiment_NILES}, also for the plateau electrons.
\label{fig:log_SMM_sim_extended_data_fig}
}
\end{figure*}

In Fig.~\ref{fig:niles-scheme} we demonstrated the influence of different optical near fields on NILES. In Extended Data Fig.~\ref{fig:log_SMM_sim_extended_data_fig}, we show the same simulations but now on a logarithmic scale and a larger energy range, including the plateau electrons.

For the homogeneous case in Extended Data Fig.~\ref{fig:log_SMM_sim_extended_data_fig}a, the direct electrons in the low energy range do not show any CEP dependence, as expected. Yet, also this panel shows the tell-tale feature of strong field physics, the plateau, like the two others (b,c). Since the peak electric field strength determines the maximum achievable energies, we observe a modulation of the highest energies as function of the CEP, leading to the well-known arches in the cut-off region~\cite{Haworth2006}. The electron energy spectrum in the homogeneous case reaches up to the expected 10 times the ponderomotive energy $U_\mathrm{P}$ of the laser field, i.e., roughly \SI{42}{\electronvolt} \cite{PaulusBeckerNicklichEtAl1994}.

This cut-off energy cannot be reached for the inhomogeneous near field profiles (Extended Data Fig.~\ref{fig:log_SMM_sim_extended_data_fig}b,c), because of a quenched quiver motion \cite{Herink2012}. This energy reduction is even better visualized by observing the final energies from one optical cycle ({\color{black_red} Extended Data} Fig.~\ref{fig:shape_of_NILES}a), where the maximum energies only reach \(8.3\,U_\mathrm{P}\) for the chosen parameters. In addition, the reduced near field leads to a delayed starting time of rescattering, indicated by the stars.

The energy reduction demonstrates by how much NILES is different: while the optical near-field affects only the energy of the rescattered electron, its presence changes the spectral structure of direct electrons completely.

Further, the simulations depicted in Extended Data Fig.~\ref{fig:log_SMM_sim_extended_data_fig}a-c show that the near field decay changes the CEP $\phi_\mathrm{max}$ at which the electrons are born that reach the highest energies (indicated by colored arrows).

To determine $\phi_\mathrm{max}$ experimentally, we recorded further spectra with a CEP step size of \SI{0.26}{\radian} over a range of over three periods with $5\times10^5$ electrons per phase step and rate of 0.4 electrons per pulse (Extended Data Fig.~\ref{fig:absolute_phase}a). We determine the cut-off position for each CEP by fitting two linear functions to the plateau and the cut-off region \cite{Krger2011,Thomas2013}.

We fit a sine to the determined cut-offs shown in Extended Data Fig.~\ref{fig:absolute_phase}a and find an experimental value of $\phi_\mathrm{max,exp}=1.8\pm 0.5$\,rad and a cut-off position of $40.7$\,eV with a modulation depth of \SI{5.4\pm0.2}{\electronvolt} (twice the amplitude of the sine fit). The resulting near field intensity in this measurement is then \SI{2.0e13}{\watt\per\square\centi\meter}.

In Extended Data Fig.~\ref{fig:absolute_phase}b we show the simulated phase $\phi_\mathrm{max}$ as function of the decay constant $\zeta$. We find that $\phi_\mathrm{max}$ monotonically decreases for increasing $\zeta$. We observe that $\phi_\mathrm{max}(\zeta)$ is matched by a power law, where the offset is given by the homogeneous field case, with a phase of $\phi_{max}(\zeta \rightarrow\infty) = \SI{0.9}{\radian}$. The measured phase of $1.8\pm 0.5$ matches the simulated data within the error bars (black dot).

Because the experimentally determined phase $\phi_\mathrm{max,exp}$ is \SI{0.9}{\radian} larger than for the simulated case with a homogeneous optical field ($\zeta = \infty$), we see that the CEP at which the highest electron energies are created cannot be considered constant in the presence of an optical near field. This result shows that schemes measuring the CEP by the cut-off electrons clearly have to take decaying near fields into account. If not, the CEP can be grossly misinterpreted.

Moreover, the evaluated modulation depth of the cut-off of \SI{5.4\pm0.2}{\electronvolt} serves as an intrinsic measure for the near field pulse duration present at the tip's apex. The modulation depth is in good approximation given by the difference of the ponderomotive energies associated to the maximum field at CEP$=0$ and CEP=$\pi$, for one fixed pulse envelope. In {\color{black_red}Extended Data} Fig.~\ref{fig:absolute_phase}c we show the modulation depth as function of the cut-off energy, i.e. ten times the ponderomotive energy, and the pulse duration. Based on this map, the experimental pulse duration is \SI{11.1\pm0.4}{\femto\second}, in good agreement to a FROG-based measurement of $11.5$\,fs. 

Similar maps can be generated using the semi-classical simulation including both the optical near field and the static field. The map produced by such a simulation gives then a pulse duration of \SI{11.5\pm0.4}{\femto\second}, perfectly matching the result from FROG. The better agreement is due to the fact that quenching effects are not taken into account when we only consider the difference of ponderomotive energies at CEP$=0$ and CEP=$\pi$. However, we emphasize that using this difference provides experimenters with a simple tool to measure the pulse duration within the typically required accuracy in the experiment, without the need for simulations.

\subsection*{Impact of coherent phenomena on NILES}
So far all simulations were based on a semi-classical treatment, where only the emission process was modeled quantum mechanically. To rigorously include quantum interference and quantum diffusion, we now simulate the energy spectra based a numerical solution of the one-dimension time-dependent Schrödinger equation (TDSE) \cite{Seiffert2018,Dienstbier2023}. We obtain the wave function of the electron by integrating the TDSE numerically by a Crank-Nicolson method (for details {\color{black_red} and code} see~\cite{Dienstbier2023}). As ground state, we assume an electron which is bound in a box potential with a half-sided linear decaying potential, mimicking the static potential present at the tip. The potential is chosen such that the electron has an energy of \SI{20}{\electronvolt} without any laser field after traversing the full potential. The width of the box is chosen to match the work function of tungsten, which we assume as 6\,eV \cite{Dienstbier2023}. {\color{black_red} The assumption of one single bound state may seem surprising because metals have in general many available states that are occupied according to the Fermi-Dirac distribution. However, because of the high non-linearity of the electron emission, one energy level will be dominant and all other lower lying states are strongly suppressed (see also \cite{Dienstbier2022_phd}}).

\begin{figure*}[h]
\includegraphics[width=0.99\linewidth]{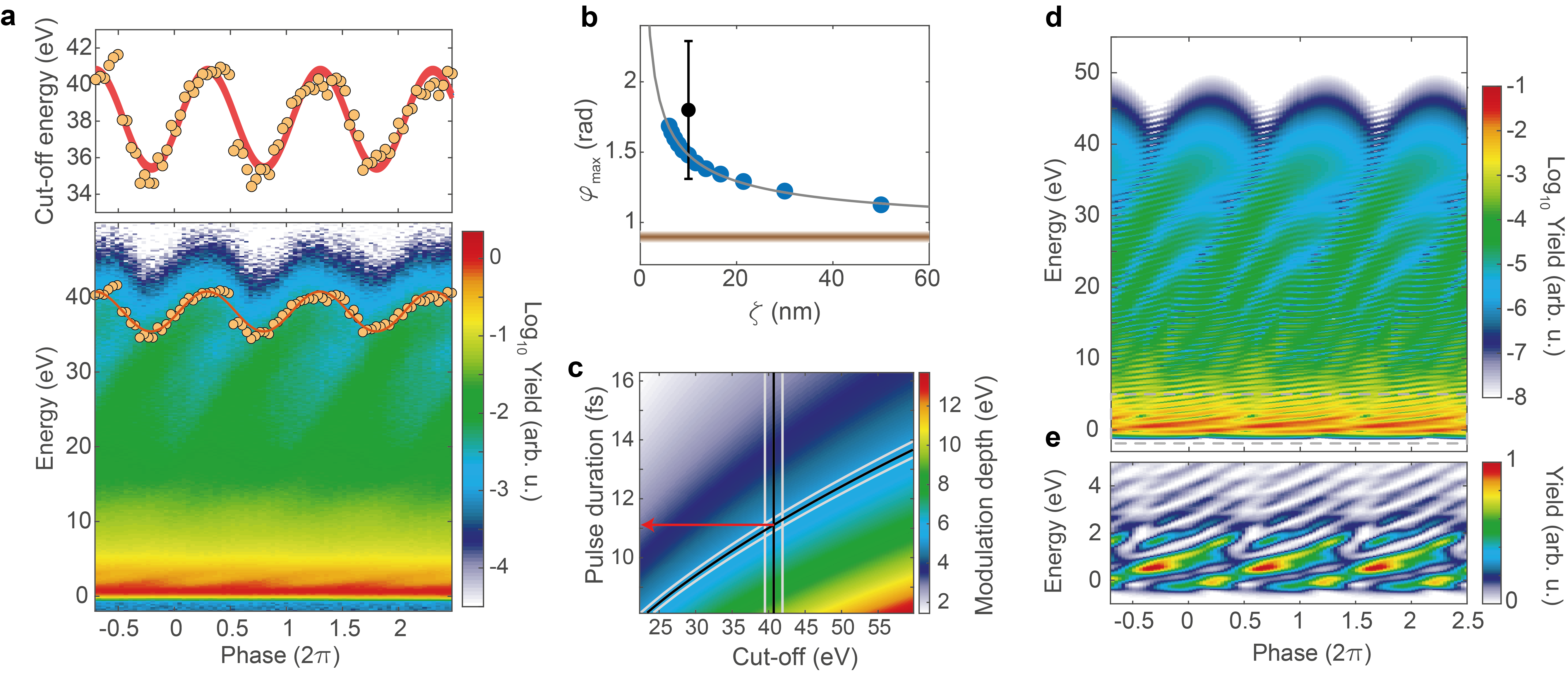}
\caption{\textbf{Plateau electrons and TDSE simulation results.} \textbf{a} Bottom: Experimental spectrum on a logarithmic scale. Orange dots are the determined cut-off positions which we fit by a sine, shown also as zoom-in in the upper panel. We note that these data were recorded with a tungsten tip with a tip radius of 10\,nm without FIB milling.  \textbf{b} Phase shift $\phi_\mathrm{max}$ as function the decay constant $\zeta$. The blue dots are the phase extracted from the semi-classical simulation, and the black dot is the experimentally determined phase shift from \textbf{a}. \textbf{c} Modulation depth of the cut-off as function of pulse duration and cut-off position, calculated from the difference of the ponderomotive energy associated with the maximum electric field at CEP$=0$ and CEP=$\pi$ (see text for details). The black lines show the measured cut-off (vertical axis) and the modulation depth (curved) from \textbf{a}. The grey curves indicate the standard deviation of the measurement. The resulting pulse duration is given by the intersection of the two black lines, shown by the red arrow.  \textbf{d} Solution of the time-dependent Schrödinger equation with parameters similar to the experiment in \textbf{a}. Because of the fully coherent simulation, interference effects like mulitphoton-peaks (MPP) appear, visible over the entire spectrum. \textbf{e} Low-energy region of \textbf{d}. Both NILES (thick features) from 0-2\,eV and MPP-peaks above 2\,eV are visible. The interference of both features leads to an additional yield modulation of NILES, while the shape is mostly unaffected. See text for details.   \label{fig:absolute_phase}}
\end{figure*}

In Extended Data Fig.~\ref{fig:absolute_phase}d we show the result for an incident intensity of \SI{1.8e13}{\watt\per\square\centi\meter}, $\zeta = \SI{10}{\nano\meter}$, and $\xi_0=5$,  similar to the experiment shown in Extended Data Fig.~\ref{fig:absolute_phase}a. {\color{black_red} The main features of the simulated spectrum are very close to the spectra simulated with the semi-classical model. However,} the TDSE is fully coherent, hence we find now multi-photon photoemission peaks (MPP) spaced by the photon energy, well visible as fine stripes over the entire spectrum. NILES also shows up in the low-energy region, smoothly transitioning into the spacing of the MPP-peaks above \SI{2}{\electronvolt} (Extended Data Fig.~\ref{fig:absolute_phase}e). It is interesting to note is that the whole low-energy region (-1 to 5\,eV) experiences a CEP-dependent energy shift similar to the shape of NILES, even above \SI{2}{\electronvolt} where we classically observe no NILES feature.
{\color{black_red}The fact that the semi-classical and the quantum simulation agree except for the appearance of MPP-peaks, shows that the dynamics of the emitted eletron wave packets is dominated by the center of mass motion in the optical near field (Ehrenfest dynamics).}

In the experiment we have so far not observed MPP-peaks together with NILES{\color{black_red}, which can be for two reasons. For longer pulse durations without CEP variation, we have measured such MPP-peaks already in the same setup \cite{Heimerl2023}. The broader bandwidth of our few-cycle laser pulses in the present case broadens the width of the MPP-peaks as is shown in \cite{Luo2021}}. {\color{black_red} Further,} the coherence time at such high intensities is {\color{black_red}possibly} reduced so heavily that already two emission events in two consecutive cycles cannot interfere anymore. Investigating details of the electron coherence will be subject of future work.  

\newpage

\begin{figure*}[h]
\includegraphics[width=\linewidth]{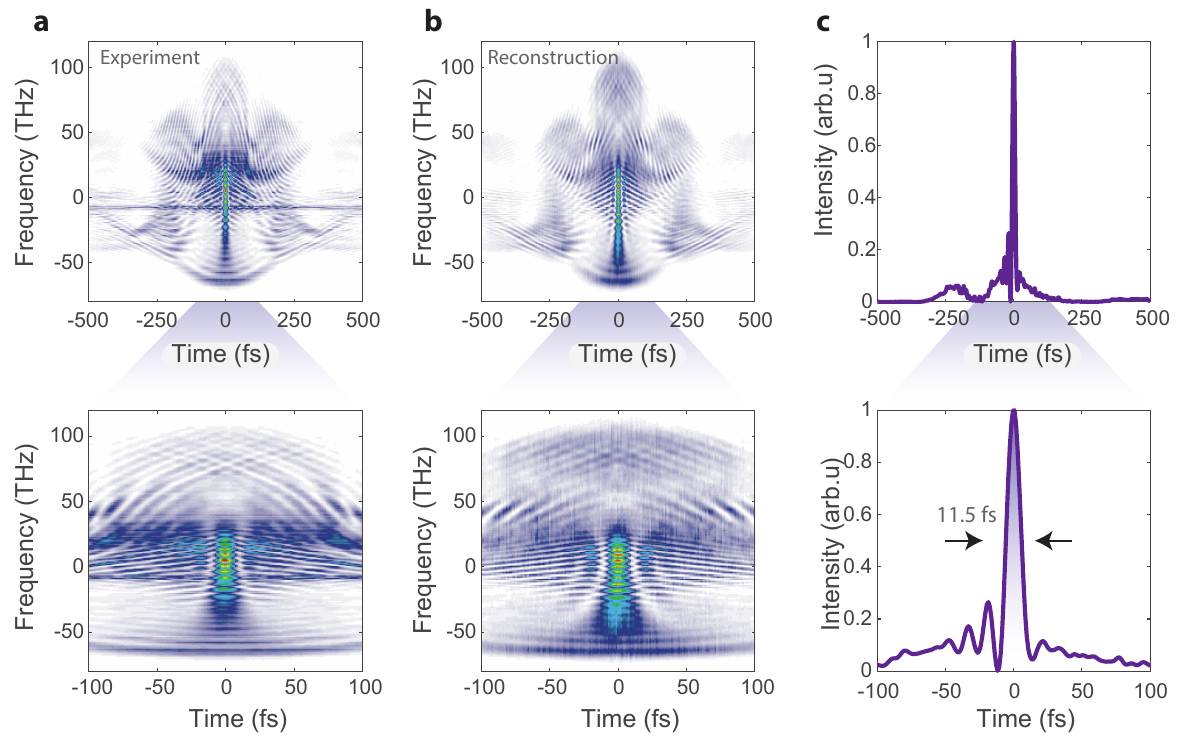}
\caption{\textbf{FROG measurement.} \textbf{a} Experimental second-harmonic FROG trace (upper panel) with close-up (lower panel). \textbf{b} Reconstructed FROG trace. \textbf{c} Retrieved temporal intensity with a $\tau_\mathrm{FWHM} = \SI{11.5}{\femto\second}$}
\label{fig:FROG_measurement}
\end{figure*}
\newpage

\begin{figure*}[h]
\includegraphics[width=\linewidth]{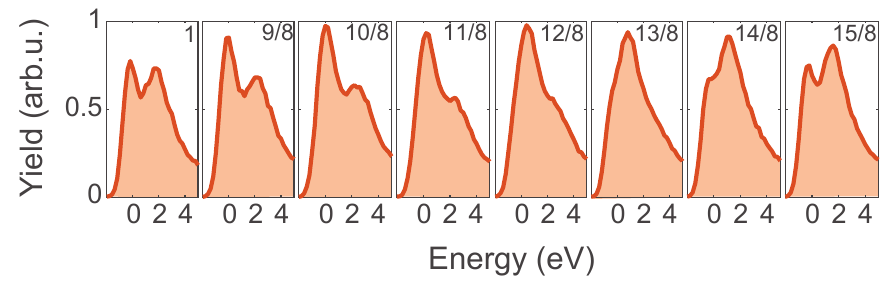}
\caption{\textbf{Line-out of experimental data}. Low-energy region of the experimental data of Fig.~\ref{fig:experiment_NILES}, where NILES shows up, visible as peaks in the line-outs. The top-right values indicate the CEP as multiples of $2\pi$.}
\label{fig:lineouts_experiment}
\end{figure*}

\newpage



\newpage

\newpage